\documentclass[preprint]{elsarticle}

\usepackage{natbib}
\usepackage[utf8]{inputenc}
\usepackage[T1]{fontenc}
\usepackage[hyphens,spaces,obeyspaces]{url}
\usepackage{booktabs}
\usepackage{tabularx}
\usepackage[colorlinks=true,allcolors=blue]{hyperref}
\usepackage{pythonhighlight}
\usepackage{multirow}
\usepackage{amsmath,amssymb,amsfonts}
\usepackage{algorithmic}
\usepackage{graphicx}
\usepackage{textcomp}
\usepackage{xcolor}
\usepackage{subcaption}
\usepackage{circledsteps}
\usepackage{rotating}

\usepackage[frozencache,cachedir=minted-cache]{minted}
\definecolor{LightGray}{gray}{0.95}

\def\BibTeX{{\rm B\kern-.05em{\sc i\kern-.025em b}\kern-.08em
    T\kern-.1667em\lower.7ex\hbox{E}\kern-.125emX}}

\usepackage{changepage}
\usepackage{framed}
\makeatletter
{\par\unskip\endMakeFramed}
\makeatother
\definecolor{formalshade}{rgb}{0.93,0.93,0.93}
\definecolor{darkblue}{rgb}{0.2, 0.2, 0.2}
\newenvironment{formal}{%
\def\FrameCommand{%
  \hspace{1pt}%
  {\color{darkblue}\vrule width 2pt}%
  {\color{formalshade}\vrule width 4pt}%
  \colorbox{formalshade}%
}%
\MakeFramed{\advance\hsize-\width\FrameRestore}%
\noindent\hspace{-1pt}
\begin{adjustwidth}{}{7pt}%
\vspace{2pt}\vspace{2pt}%
}
{%
\vspace{3pt}\end{adjustwidth}\endMakeFramed%
}
\newcounter{resultcounter}

\newcounter{patterncounter}

\newcommand{\bc}[1]{DABC#1}
\newcommand{\lbc}[1]{\textsc{Default Argument Breaking Change#1}}
\newcommand{\scikit}{\textsc{Scikit-Learn}}
\newcommand{\pandas}{\textsc{Pandas}}
\newcommand{\numpy}{\textsc{NumPy}}
\newcommand{\bcircle}[1]{\Circled[fill color=black, inner color=white]{#1}}
\newcommand{\gcircle}[1]{\Circled[fill color=LightGray, outer color=gray, inner color=black]{#1}}

\newcommand{\rqA}{RQ.1: What are the Most Common \bc{s}?~}
\newcommand{\rqB}{RQ.2: In which Version the \bc{s} were Introduced?~}
\newcommand{\rqC}{RQ.3: In which Modules the \bc{s} were Introduced?~}

\newcommand{\rqD}{RQ.4: How Clients are Vulnerable to \bc{s}?}

\newcommand{\rqE}{RQ.5: Why API maintainers introduce \bc{s}?}
\newcommand{\rqF}{RQ.6: Which effects do \bc{s} have on API clients?}

\journal{Journal of Systems and Software}
\begin{document}
\begin{frontmatter}
    \title{Unboxing Default Argument Breaking Changes in 1 + 2 Data Science Libraries}

    \author[inst1]{João Eduardo Montandon\corref{cor1}}
    \ead{joao@dcc.ufmg.br}
    \author[inst2]{Luciana Lourdes Silva}
    \ead{luciana.lourdes.silva@ifmg.edu.br}
    \author[inst3]{Cristiano Politowski}
    \ead{cristiano.politowski@ontariotechu.ca}
    \author[inst1]{Daniel Prates}
    \ead{danielprates@ufmg.br}
    \author[inst1]{Arthur de Brito Bonifácio}
    \ead{arthurb@ufmg.br}
    \author[inst4]{Ghizlane El Boussaidi}
    \ead{ghizlane.elboussaidi@etsmtl.ca}

    \cortext[cor1]{Corresponding Author}

    \affiliation[inst1]{
        organization={Universidade Federal de Minas Gerais},
        city={Belo Horizonte},
        country={Brazil}
    }
    \affiliation[inst2]{
        organization={Instituto Federal de Minas Gerais},
        city={Ouro Branco},
        country={Brazil}
    }
    \affiliation[inst3]{
        organization={Ontario Tech University},
        city={Oshawa},
        country={Canada}
    }
    \affiliation[inst4]{
        organization={École de Technologie Supérieure},
        city={Montréal},
        country={Canada}
    }

    \begin{abstract}
        Data Science (DS) has become a cornerstone for modern software, enabling data-driven decisions to improve companies services.
        Following modern software development practices, data scientists use third-party libraries to support their tasks.
        As the APIs provided by these tools often require an extensive list of arguments to be set up, data scientists rely on default values to simplify their usage.
        It turns out that these default values can change over time, leading to a specific type of breaking change, defined as Default Argument Breaking Change (DABC).
        This work reveals 93 DABCs in three Python libraries frequently used in Data Science tasks---Scikit Learn, NumPy, and Pandas---studying their potential impact on more than 500K client applications.
        We find out that the occurrence of DABCs varies significantly depending on the library;
        35\% of Scikit Learn clients are affected, while only 0.13\% of NumPy clients are impacted.
        The main reason for introducing DABCs is to enhance API maintainability, but they often change the function's behavior.
        We discuss the importance of managing DABCs in third-party DS libraries and provide insights for developers to mitigate the potential impact of these changes in their applications.
    \end{abstract}



\end{frontmatter}

\section{Introduction}

Data Science (DS) is the process of extracting knowledge from data, enabling organizations to make informed decisions~\cite{Provost2013, Ramasamy2022, Quaranta2021, Quaranta2022}.
Due to the increasing data availability, it has become a cornerstone for modern software products, allowing companies to make data-driven decisions to improve their services.
Companies like Netflix, Google, and Microsoft keep dedicated teams of data scientists who build and maintain many predictive models, used to recommend movies to watch, filter web pages to see, and perform real-time text translation~\cite{NetflixTechnologyBlog2019, Winters2020, Amershi2019}.

Typically, the DS pipeline involves several steps; from data collection and analysis, to model training, evaluation, and deployment~\cite{Provost2013, Amershi2019}.
Following the practices of modern software development, data scientists also rely on third-party components to perform these tasks~\cite{Montandon2019b}.
In this context, the Python ecosystem stands out from others due to the options available~\cite{Haryono2021, Grotov2022, Zhang2021, Montandon2021, Montandon2021a}, including libraries for data analysis (e.g.,~Pandas), model training (e.g.,~Scikit Learn), and data visualization (e.g.,~Matplotlib).

These tools provide comprehensive APIs for developers interested in using their features.
However, some of these APIs are complex, requiring a significant number of arguments to be set up~\cite{Zhang2021}.
Considering the Scikit Learn API as an example, the constructor of \mintinline{python}|SVC|\footnote{\url{https://scikit-learn.org/stable/modules/generated/sklearn.svm.SVC.html}}---a widely used classifier based on SVM---provides 14 arguments to be defined.
Similarly, the \mintinline{python3}|concat()|\footnote{\url{https://pandas.pydata.org/pandas-docs/version/2.2/reference/api/pandas.concat.html}} function provided by Pandas requires 10 arguments.
Fortunately, the maintainers of both libraries defined a set of default values for such arguments so developers can effortlessly use these functions.
In practice, no argument is needed to create a \mintinline{python3}|SVC| model, and only one argument is necessary to call the \mintinline{python3}|concat| function.

On the other hand, relying on these default arguments increases the coupling between the library and its client applications;
now clients depend not only on the method syntax provided in the API but also on the values assigned to the arguments to keep running as expected.
Consequently, changing these values affects the client's behaviour.
For instance, Scikit Learn maintainers changed the Kernel coefficient formula used by default in \mintinline{python}|SVC|---i.e.,~\mintinline{python}|gamma| argument---from \mintinline{python}|"auto"| to \mintinline{python}|"scale"| between versions 0.21 and 0.22, which can drastically change the model's results.

The lack of backward compatibility between library versions is known as \textit{breaking change}~\cite{Brito2020, Ponomarenko2012, Venturini2023, Ochoa2022}.
Most studies focus on syntactical breaking changes~\cite{Brito2020, Ochoa2022}, which deal with changes that result in syntax errors.
On the other hand, breaking changes can also include semantical modifications~\cite{Ponomarenko2012, Mezzetti2018}, meaning that they may alter the behavior of a library without causing any syntax errors in client code.
This can be a significant issue for developers, as the changes may go unnoticed and can impact the reproducibility of their code.
We interpret this problem as a semantical variant of the breaking change problem, which we named \lbc\ (\bc).
To the best of our knowledge, we did not find any study investigating this type of breaking change.

\lbc{s} play a particular role in the context of data science tools.
First, most components provided by these libraries are hard to inspect, i.e.,~machine learning engineers may spend significant effort debugging why a given model returned a given result~\cite{Ribeiro2016a}.
\bc{s} add another difficulty layer to this process as a subtle change in a model's argument value may drastically modify its outcome.
Second, numerous Jupyter Notebooks---the tool of choice for many data scientists---lack configuration files declaring their module dependencies~\cite{Pimentel2019, Wang2020a}.
A recent study by Pimentel et al.~\cite{Pimentel2019} indicates that less than 14\% of public notebooks declare some dependency file, e.g., \textit{requirements.txt}.
In other words, most notebooks rely on the default argument values of the libraries they use.

In this paper, we investigate the occurrence of \bc{s} on three of the most popular Python libraries used in Data Science: Scikit Learn, NumPy, and Pandas.
Specifically, we identified the \bc{s} introduced in each library and analyzed their potential impact on 800K client applications.
Furthermore, we manually inspected the commit information that introduced each \bc{} to understand the reasons that motivated library maintainers to do so.
The research reported in this paper is organized into two major work units.

\paragraph{\textbf{First Work}}
We started by investigating the occurrence of \bc{s} in Scikit Learn, one of Python's most used libraries to build machine learning models.
We manually analyzed the changes made to the default values of the arguments in Scikit Learn functions---according to the official documentation---to understand the main characteristics of \bc{s}.
Then, we quantitatively investigated the impact that \bc{s} have on Scikit Learn's client applications by analyzing its use in 194,099 Jupyter Notebooks, publicly available on GitHub.
Next, we replicated the initial study with two other Python libraries: NumPy and Pandas, commonly used for data analysis and scientific computation.
This study is based on four research questions from our previous work~\cite{Montandon2023}, describing \textit{what} \bc{s} are (RQ.1), \textit{when} and \textit{where} they were introduced (RQ.2 and RQ.3), and \textit{how} clients are vulnerable to them (RQ.4).

\elsparagraph{\rqA}
We identified 77 \bc{s} declared in Scikit Learn.
By contrast, 5 and 11 \bc{s} were detected for both NumPy and Pandas, respectively.
While 56 \bc{s} point to class constructors in Scikit Learn, e.g.,~\mintinline{python}|SVC.__init__()|, only one was detected among NumPy and Pandas's \bc{s}.

\elsparagraph{\rqB}
We identified \bc{s} in eight major versions for Scikit Learn.
Version 0.22 stands out with 43 occurrences, followed by 0.20 and 1.1 with 11 each;
these three versions concentrate 84\% of all \bc{s}.
For NumPy, three \bc{s} were introduced in one minor version to tackle a security issue.
On the other hand, introducing \bc{s} in minor releases is a more common practice in Pandas;
five out of six versions are minor ones.

\elsparagraph{\rqC}
In Scikit Learn, the majority of \bc{s} were reported on \textit{Model Training} and \textit{Model Evaluation} APIs;
together, these modules hold 61 out of 77 \bc{s} (79\%).
This suggests that ML models are in the spotlight regarding \bc{s}.
\textit{DataFrame} was the most vulnerable module in Pandas, while most of NumPy's \bc{s} were found in \textit{General Functions}.

\elsparagraph{\rqD}
In Scikit Learn, 35\% client applications are vulnerable to at least one \bc{}.
Despite having a higher number of clients, we detected fewer vulnerabilities for NumPy and Pandas.
21\% of Pandas clients are vulnerable, and only 0.1\% of NumPy clients are exposed.
The presence of \bc{s} does not correlate with other software metrics, such as LOC, function coupling, and cyclomatic complexity.

\paragraph{\textbf{Extended Analysis}}
We also expanded our initial work with two new research questions to understand \textit{why} library maintainers introduce these breaking changes.
The major results of this work are as follows.

\elsparagraph{\rqE}
While \textit{Maintainability}-based changes are the most frequent ones for Scikit Learn (81\%), \textit{API Compatibility} and \textit{Bug Fixing} are the most prevalent reasons for NumPy (60\%) and Pandas (44\%), respectively.
By contrast, only two \bc{s} were introduced as a consequence of \textit{New Features}.

\elsparagraph{\rqF}
Overall, the \bc{s} we identified can lead to four distinct effects on client applications: \textit{Aesthetics}, \textit{Behavior}, \textit{Performance}, and \textit{Refactoring}.
\textit{Behavior}-based \bc{s} stand out as the most frequent ones for Scikit Learn (58; 75\%), Pandas (5; 46\%), and NumPy (4; 80\%).
In particular, only Scikit Learn has introduced \bc{s} resulting from refactoring actions, suggesting its maintainers adopt refactoring to keep the API compatible with clients' applications.

\paragraph{\textbf{Contributions}}
We summarize the contributions of this paper as follows.

\begin{itemize}
    \item We characterize an unexplored kind of semantic breaking change focused on the default values of API arguments, called \lbc\ (\bc).
    \item We study the characteristics of \bc{s} in three popular Python libraries used in Data Science, and analyze how real-world client applications are exposed to them.
    \item We qualitatively investigate why these changes were introduced by the library maintainers, and which kind of impact they potentially bring to clients.
    \item Finally, we discuss strategies that client developers should adopt to avoid being affected by \bc{s}.
\end{itemize}

\paragraph{\textbf{Paper Structure}}
This paper is organized as follows.
Section \ref{sec:background} defines in detail what a \lbc\ is, and how it can make client applications vulnerable.
Sections \ref{sec:method-collection} and \ref{sec:method-rq} describe the initial approach we adopted to analyze the \bc{s} detected in Scikit Learn, NumPy, and Pandas;
the results regarding the first four Research Questions are in Section \ref{sec:results}.
Section \ref{sec:extended-analysis} presents the method used to answer the new proposed questions and their results.
Section \ref{sec:discussion} discusses the implications of this work.
Section \ref{sec:threats} reports threats to validity, and Section \ref{sec:related-work} summarizes the related work.
Finally, we conclude this paper in Section \ref{sec:conclusion}.

\section{Background}
\label{sec:background}

\subsection{Default Arguments in a Nutshell}

The Python language supports functions that, once implemented, can be called with only some arguments.
For instance, the  \mintinline{python}|round(number, digits)| function\footnote{\url{https://docs.python.org/3/library/functions.html\#round}} has two parameters\footnote{We use arguments and parameters interchangeably in this paper.} and returns the \mintinline{python}|number| rounded to  \mintinline{python}|digits| decimal places.
It turns out that  \mintinline{python}|round()| can be called with both one and two arguments; if  \mintinline{python}|digits| is omitted, the function rounds \mintinline{python}|number| to its nearest integer value.
This means that calling \mintinline{python}|round(3.1415, 2)| returns $3.14$, while calling \mintinline{python}|round(3.1415)| gives $3.0$.
Such behavior is possible due to Default Arguments,\footnote{\url{https://docs.python.org/3/tutorial/controlflow.html\#default-argument-values}} which
specify values that functions will use if the caller provides no value to its corresponding argument.
In the above example, the function automatically assumed \mintinline{python}|digits=0| when invoking \mintinline{python}|round(3.1415)|.

One assigns a Default Argument during function definition by attributing an arbitrary value to the arguments the developer wants to become optional, as shown in Figure \ref{lst:dav}.
In this scenario, executing \mintinline{python}|sum(10, 20)| will return $30$ (\mintinline{python}|a=10|, \mintinline{python}|b=20|), \mintinline{python}|sum(10)| will return $10$ (\mintinline{python}|a=10|, \mintinline{python}|b=0|), and calling \mintinline{python}|sum()| will return $0$ (\mintinline{python}|a=0|, \mintinline{python}|b=0|).

\begin{figure}[ht!]
    \begin{minted}[
        xleftmargin=5pt,
        frame=lines,
        framesep=2mm,
        framerule=0.5pt,
        fontsize=\small,
        numbers=left,
        numbersep=5pt,
        autogobble
    ]{python}
        def sum(a=0, b=0):
            return a + b
    \end{minted}
    \caption{Function definition with Default Argument Values.}
    \label{lst:dav}
\end{figure}

Default Arguments are a powerful resource as they mimic method overloading---methods with the same name but different parameters---which is not supported in Python by default~\cite{Phillips2018}.
This concept promotes flexibility, readability, and reusability to classes interface; relevant to successful APIs~\cite{Bloch2017, Lehman1979}.

\subsection{The Scikit-Learn Library}

\scikit\ is a free and open-source machine learning library for Python, released in 2011~\cite{Pedregosa2011}.
The library implements well-known supervised and unsupervised machine learning algorithms, such as linear and logistic regressions, support vector machines, decision trees, and k-means clustering.
The library also provides techniques to manage, evaluate, and deploy the above-mentioned models.

Such features contributed to its adoption in several industry and research projects.
\scikit\ is one of the most popular machine learning libraries worldwide, with more than 1 million downloads daily.\footnote{According to \url{https://pypistats.org/}}
As of Jan 24th, 2023, the \scikit\ repository on GitHub has more than 52,7K stars, almost 30K commits, and was forked above 23,9K times.

Despite its wide adoption, \scikit{}'s API is constantly changing.
For instance, version 1.0 was released in September 2021; the maintainers followed with eight new versions since then.
This scenario may challenge the developers whose client applications depend on \scikit\ features.

\paragraph{Default Arguments in Scikit-Learn}
\scikit{}'s API extensively relies on Default Arguments
so users can set up the models available in the library with low effort.
For example, the \mintinline{python}|SVC| class provides 14 parameters to be defined through its constructor.\footnote{\url{https://scikit-learn.org/1.1/modules/generated/sklearn.svm.SVC.html}}
These arguments are responsible for configuring several aspects of a SVC model, including the Kernel type used by the model (\mintinline{python}|kernel|), its Kernel coefficient (\mintinline{python}|gamma|), random seed values (\mintinline{python}|random_state|), etc.

Since all arguments have default values assigned, the user can quickly get started without defining any parameter to the model.
Figure \ref{lst:dav-sklearn} exemplifies this fact when creating a SVC model.
Except for \mintinline{python}|random_state|, all arguments rely on default values; \mintinline{python}|kernel| was defined to \mintinline{python}|"rbf"|, and \mintinline{python}|gamma| was assigned to \mintinline{python}|"scale"|.\footnote{Default values available from version 1.1.2, the latest stable one at the time of this work.}

\begin{figure}[ht!]
    \begin{minted}[
        xleftmargin=5pt,
        frame=lines,
        framesep=2mm,
        framerule=0.5pt,
        fontsize=\small,
        numbers=left,
        numbersep=5pt,
        autogobble
    ]{python}
        from sklearn.svm import SVC

        clf = SVC(random_state=42)
        clf.fit(X_train, y_train)
    \end{minted}
    \caption{Default Argument Values in action in \scikit.}
    \label{lst:dav-sklearn}
\end{figure}

\subsection{NumPy and Pandas Libraries}

In our extended work, we included two additional data science libraries: \numpy\ and \pandas.
These libraries were selected based on their popularity as indicated by the 2022 Stack Overflow Developer Survey.\footnote{\url{https://survey.stackoverflow.co/2022/\#section-most-popular-technologies-other-frameworks-and-libraries}}
According to the survey, all three libraries are among the top-10 most used libraries in the Data Science field;
\scikit\ is used by 12.6\%,  \numpy\ by 28.7\%, and \pandas\ by 25.1\% of the respondents.
Besides, all three projects have very active communities, as summarized in Table \ref{tab:libs-stats}.

\begin{table}[ht]
    \small
    \centering
    \begin{tabularx}{1\linewidth}{lXlrrr}
        \toprule
        \textbf{Library} & \textbf{Short Bio}                  & \textbf{Release} & \textbf{\#Stars} & \textbf{\#Commits} \\ \midrule
        \scikit\         & Module for machine learning tasks.  & 2011             & 52.7K            & 30K                \\
        \numpy\          & Package for scientific computing.   & 2005             & 25K              & 35K                \\
        \pandas\         & Data analysis/manipulation library. & 2009             & 40K              & 33K                \\
        \bottomrule
    \end{tabularx}
    \caption{The three libraries studied in this work.}
    \label{tab:libs-stats}
\end{table}

\paragraph{Pandas}
Released in 2009, \pandas\ is a data analysis and manipulation library.\footnote{\url{https://pandas.pydata.org/about/}}
At its core, the library provides highly efficient data structures for working with structured data---such as Data Frames and Series---and tools for cleaning, reshaping, and combining the information stored in these structures.
Such features make it widely adopted in Data Science and Machine Learning projects, with more than 5M downloads on a daily-basis.\footnote{\url{https://pypistats.org/packages/pandas}}
The \pandas\ library is also one of the most active projects on GitHub; it contains more than 33K commits, 40K starts, and 16.9K forks.

\paragraph{NumPy}
Launched in 2005, \numpy\ stands as the major Python library for numerical computing.
Designed to handle arrays and matrix manipulation efficiently, \numpy\ comes with tools for several mathematical operations and analysis.
The library is widely used in various scientific and engineering fields, serving as a fundamental tool for tasks like linear algebra and statistical analysis.
Widely used for scientific, engineering, and machine learning tasks, \numpy\ has more than 7M downloads, daily.\footnote{\url{https://pypistats.org/packages/numpy}}
Its repository also features among the popular ones in Python's ecosystem, with more than 25K starts and 8K forks.\footnote{GitHub repository: \url{https://github.com/numpy/numpy}}

\subsection{What is a Default Argument Breaking Change (\bc)?}

Despite the advantages of using Default Arguments, they might bring issues to client applications relying on them.
Specifically, library maintainers can update the default values of some parameters to meet new conditions.
Changes of this nature do not break clients' code since the function signature (name and arguments) remains the same.
Nevertheless, they might introduce incompatibilities as the new value changes the function's behaviour.

For instance, \scikit\ maintainers updated the value of \mintinline{python}|gamma| argument of the \mintinline{python}|SVC| classifier from  \mintinline{python}|"auto"| to  \mintinline{python}|"scale"| in version $0.22$.
This update clearly affects models relying on this default value as it changes the math formula used to calculate the \textit{gamma} value.
Consequently, the code that creates and trains a SVC model in Figure \ref{lst:dav-sklearn} outputs very different results in versions $0.21$ and $0.22$.

\paragraph{Default Argument Breaking Change Example}
We illustrate this maintenance problem by implementing the minimum working example in Figure \ref{lst:dabc-example}.
This example classifies the 20 newsgroup dataset, a popular real-world collection containing 18,000 newsgroup posts grouped into 20 distinct topics.
The Stanford Natural Language Processing Group collected this dataset over several years, and it has become a popular alternative for experiments in text applications of machine learning techniques.
Currently, it has been used as a benchmark in popular research works~\cite{Wang2016, Bianchi2021}.

\begin{figure}[ht!]
    \begin{minted}[
        xleftmargin=5pt,
        frame=lines,
        framesep=2mm,
        framerule=0.5pt,
        fontsize=\small,
        numbers=left,
        numbersep=5pt,
        autogobble
    ]{python}
        from sklearn import datasets
        from sklearn.model_selection import train_test_split
        from sklearn.metrics import accuracy_score
        from sklearn.svm import SVC
            
        # Load dataset
        ds = datasets.fetch_20newsgroups_vectorized()
        X = ds.data[:, 2:]
        y = ds.target
            
        # Create training/test data split
        X_train, X_test, y_train, y_test = train_test_split(X, y, test_size=0.3, 
                                            random_state=42, stratify=y)
            
        # Create an instance of SVC Classifier 
        clf = SVC(random_state=42)

        # Fit, predict, and measure model's performance
        clf.fit(X_train, y_train)
        y_pred = clf.predict(X_test)
        print('Acc: %.3f' % accuracy_score(y_test, y_pred))
    \end{minted}
    \caption{Illustrative example of \bc\ in \scikit.}
    \label{lst:dabc-example}
\end{figure}

In this script, we first download the 20 newsgroup dataset and obtain their descriptive and predictive variables (lines 7--9).
Next, as with typical ML applications, we split the data into training and test groups in lines 11--16.
In line 19, we create a new \mintinline{python}|SVC| instance; we intentionally did not define any argument except for \mintinline{python}|random_state| to ensure the same randomness will be present in any execution.
Finally, lines 21--24 fit the model with the training data, predict it with test data, and measure its accuracy level.

We execute this script using \scikit\ in both versions 0.21 (\mintinline{python}|gamma="auto"|) and 0.22 (\mintinline{python}|gamma="scale"|).
In version 0.21, we scored 0.05 points for accuracy.
The same script reached 0.82 points for accuracy in version 0.22, a difference of 77 points.

This issue encompasses a specific type of breaking change, which we named \lbc\ (\bc).
This paper focuses on characterizing \bc{s} in three Data Science libraries and measuring its impact on client applications.

\section{Data Collection}
\label{sec:method-collection}

All three projects adopt strict contribution guides to enforce better software practices, with source code documentation conventions included.
For example, the \scikit\ project provides specific instructions to report changes in ``\textit{the default value of a parameter}''.\footnote{The contribution guide can be accessed at \url{https://scikit-learn.org/stable/developers/contributing.html}.}
According to these guidelines, every modification involving the value of an argument should have its docstring's documentation annotated with the $versionchanged$ directive.
Also, the old and new default values should be reported together with the version the change became effective.

Figure \ref{lst:versionchanged-example} presents an example of this description in \scikit\ project.
In this case, the parameter \mintinline{python}|gamma|---which belongs to the constructor of \mintinline{python}|SVC| class---had its value changed from \mintinline{python}|"auto"| to \mintinline{python}|"scale"|, valid from version 0.22 onward.
We relied on this guideline to collect the \lbc{s} studied in this work.

\begin{figure}[ht!]
    \begin{minted}[
        frame=lines,
        framesep=2mm,
        framerule=0.5pt,
        fontsize=\small,
        breaklines,
        autogobble
    ]{python}
        class SVC(BaseSVC):
            """C-Support Vector Classification.
            ...
            Parameters
            ----------
            ...
            gamma : {'scale', 'auto'} or float, default='scale'
                Kernel coefficient for 'rbf', 'poly' and 'sigmoid'.
                - if ``gamma='scale'`` (default) ...
                - if 'auto', ...
                - if float, ...
                .. versionchanged:: 0.22
                The default value of ``gamma`` changed from 'auto' to 'scale'.
            """
    \end{minted}
    \caption{Example of a default argument changed in \scikit\ documentation.}
    \label{lst:versionchanged-example}
\end{figure}

\paragraph{Mining Changes on Functions Arguments}
On October 11th, 2022, we cloned the \scikit\ project from GitHub,\footnote{Repository available at \url{https://github.com/scikit-learn/scikit-learn/}.} and manually checked out the commit of version 1.1.2; the latest public release available.
Next, we selected all Python files in the $sklearn$ directory, as the library's source files are located in this directory.
We then opened each Python file and filtered out the lines containing the $versionchanged$ directive;
we did this using the regular expression ``$\backslash.\backslash. versionchanged:: .+$''.
This procedure initially returned 179 occurrences.

On April, 9th, 2023, one author has cloned the \pandas\ project and checked out the version 2.0.0.
From the \textit{pandas} directory, he looked for the \textit{versionchanged} directive using the same regular expression, detecting 126 \textit{versionchanged} occurrences in total.
As for \numpy, another author cloned and checked out version 1.24.3 on June, 2nd, 2023.
He collected 54 \textit{versionchanged} occurrences from Python files inside \numpy's source folder.

\paragraph{Selected Attributes}
For each occurrence, we collected a list of five attributes associated to each occurrence.
These attributes were used to answer the re\-search questions proposed both in our initial study (Section \ref{sec:method-rq}) and extended analysis (Section \ref{sec:extended-analysis}).
We listed these attributes below:

\begin{itemize}
    \item \textit{dabc\_msg} (RQ.1 and RQ.6): This attribute contains the message used to justify each occurrence.
          We collected this information manually for each occurrence during the categorization we did to answer RQ.1.
          Later, this information was used to answer RQ.6 in our extended analysis.

    \item \textit{version} (RQ.2): This attribute keeps the version assigned to each occurrence as collected by the regex matching procedure.
          We used this information to answer RQ.2.

    \item \textit{path} (RQ.3): Extracted during the repository analysis, this attribute specifies the relative path from each library source directory to the file where the \bc\ is located.
          We used this attribute to leverage the modules to answer RQ.3.

    \item \textit{fqn} (RQ.1 and RQ.4): This attribute holds the full qualified name of the function where the occurrence was found, i.e.,~class name followed by the method signature.
          We used this information to identify the functions and arguments affected by \bc{s} to answer RQ.1 and RQ.4.

    \item \textit{dabc\_url} (RQ.3, RQ.4, and RQ.5): We generate the GitHub URL referring to the exact point in the source code where the \bc\ was declared.
          We refer to this URL whenever we need more context about the \bc\ to answer the research questions, especially RQ.3 and RQ.4.
          Later, we accessed this URL to obtain the commit message and comments needed to answer RQ.5.
\end{itemize}

\section{Research Questions}
\label{sec:method-rq}

In this section, we describe the methodology steps to answer the four research questions proposed in the initial study.

\subsection{\rqA}
\label{sec:method-rqa}

This RQ aims to identify the \lbc{s} from the changes in arguments retrieved in Section \ref{sec:method-collection}.
For our first experiment with \scikit, three authors manually inspected and discussed all 179 occurrences, selecting the ones they considered valid.
For this, they relied on a two-step filtering approach.

First, the three authors read the description below each $versionchanged$ to select the occurrences dealing only with arguments, i.e.,~they filtered out unrelated changes.
The authors discarded 46 occurrences in this step reporting other changes, such as return values, object attributes, function refactorings, etc.
Then, they analyzed each of the remaining 133 occurrences in detail to verify which of the changes can be characterized as \lbc{}.
They discarded another 56 occurrences performing other changes in argument values, such as type changes (e.g.,~the \mintinline{python}|min_samples_split| parameter started accepting float values in version 0.18),\footnote{\url{https://scikit-learn.org/stable/modules/generated/sklearn.ensemble.RandomForestClassifier.html}} and in other values that can be passed to the parameter besides the default one (e.g.,~the \mintinline{python}|metric| argument no longer accepts a specific value in version 0.19).\footnote{\url{https://scikit-learn.org/stable/modules/generated/sklearn.neighbors.NearestCentroid.html}}
After removing these cases, we obtained 77 \bc{s} for \scikit.

We repeated this procedure for \numpy\ and \pandas.
Two authors manually inspected the 126 occurrences in \pandas\ and discarded 115 of them, resulting in 11 \bc{s}.
For \numpy, other two authors analyzed the 54 occurrences and discarded 49, leading to five \bc{s}.

\subsection{\rqB}
\label{sec:method-rqb}

This RQ aims to discover in which part of the release cycle these changes are introduced.
To do so, we analyzed the distribution of \bc{s} among the versions released by the libraries.

\paragraph{\scikit}
As described in \scikit\ documentation,\footnote{\url{https://scikit-learn.org/stable/developers/maintainer.html\#releasing}} the project maintainers nominate its releases based on PEP101.
This specification describes the library's versions using the $X.Y.Z$ triplet.
Minor versions are tracked by the $.Z$ suffix and should include bug fixes and some relevant documentation changes only, i.e.,~they should not contain a behaviour change besides a bug fix.
On the other hand, major versions indicating new releases are annotated with the $X.Y$ prefix; these versions can contain new features and significant maintenance that modify the library behaviour.

We leveraged all versions containing the $X.Y.Z$ syntax released before version 1.1.2 from the git tags available in the \scikit\ git repository.
Next, we annotated them as $major$ or $minor$ according to the \scikit\ convention described above.
We identified 56 versions released in a 12-year period, where 26 are major and 30 are minor versions.
Finally, we triangulated the release information with the $version$ field of each \bc{} identified in Section \ref{sec:method-collection}.
For example, the \bc\ present in Figure \ref{lst:versionchanged-example} matches with version 0.22, so we considered the maintainers introduced this \bc\ in a major release.

\paragraph{\numpy\ and \pandas}
Both \pandas\ and \numpy\ use different versioning strategies when compared to \scikit.
\pandas\ started adopting a loose variant of semantic versioning since version 1.0.0\footnote{\url{https://pandas.pydata.org/docs/whatsnew/v1.0.0.html\#new-deprecation-policy}} to control the evolution of its API.
The library follows the \textit{major.minor.patch} triplet to generate its releases.\footnote{\url{https://pandas.pydata.org/docs/development/policies.html\#policies-version}}
According to their documentation, API-breaking changes should occur only in major releases.
Likewise, \numpy\ relies on versioning triplet composed by \textit{major.minor.bugfix}.\footnote{\url{https://numpy.org/doc/stable/dev/depending_on_numpy.html}}
However, as it does not follow semantic versioning, API-breaking changes can be introduced in minor releases, as long as they are preceded by deprecation warnings for at least two releases.\footnote{\url{https://numpy.org/neps/nep-0023-backwards-compatibility.html}}

\subsection{\rqC}
\label{sec:method-rqc}

This RQ aims to investigate how the \bc{s} are distributed among the modules of the libraries.
We adopted two distinct strategies to answer this question, one for \scikit\ and another for both \numpy\ and \pandas.

\paragraph{\scikit}
We named nine modules that comprise the phases of Machine Learning~\cite{Domingos2012}.
For each commit, we manually inspected the changed source code files using \textit{path} and  \textit{dabc\_url} (described in Section~\ref{sec:method-collection}). Then, these changes were labelled into one of the following modules:
\begin{itemize}
    \item \textit{Dataset} contains utilities to handle large datasets (e.g., functions to download and load data) and traditional datasets (e.g., load and get data from a public repository).
    \item \textit{Data preprocessing} comprises utility functions and transformation techniques to apply on raw features for standardizing datasets.
    \item \textit{Data Decomposition} consists of functions that implement dimensionality reduction or feature selection techniques to apply to the dataset.
    \item \textit{Data Analysis} contains the implementation of statistical techniques to support the understanding of data process.
    \item \textit{Feature Processing} includes techniques to transform arbitrary data into usable data supported by Machine Learning algorithms. 
    \item \textit{Model Training} consists of algorithms' implementations for unsupervised and supervised learning methods.
    \item \textit{Model Evaluation} contains several techniques to measure the estimator performance and evaluate the model predictions' quality.
    \item \textit{Utils} comprises several utilities, such as estimate class weights for unbalanced datasets.
    \item \textit{Pipeline} consists of utilities to build a composite estimator. We followed \scikit's documentation to label this module.
    \item \textit{Others} contains changes that do not fit into any of the previous categories, e.g., exception handlers.
\end{itemize}

\paragraph{\numpy\ and \pandas}
Unlike \scikit, the features provided by both libraries do not require a strict sequence to be used.
Therefore, we manually examined their physical structure to determine their modules.
For \pandas, we looked at the official documentation\footnote{\url{https://pandas.pydata.org/docs/reference/index.html}} of each \bc's function and defined its module based on the top-level package provided by the API.
For example, one \pandas's \bc{} was assigned to the \textit{Series} module, which is the top-level package informed by the API documentation.\footnote{\url{https://pandas.pydata.org/pandas-docs/version/2.0/reference/api/pandas.Series.between.html}}
Similarly, we manually analyzed \numpy's official documentation and categorized its submodules based on the structure reported there, e.g., one \bc{} belongs to the \textit{Linear Algebra} submodule.\footnote{\url{https://numpy.org/doc/1.24/reference/generated/numpy.linalg.lstsq.html\#numpy.linalg.lstsq}}

\subsection{\rqD}
\label{sec:method-rqd}

\begin{figure*}[t!]
    \centering
    \includegraphics[width=1\linewidth]{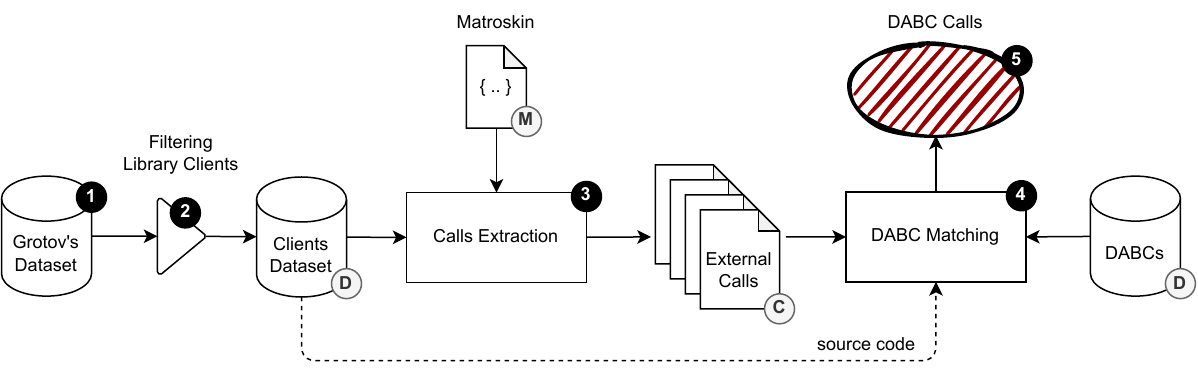}
    \caption{The data-collection pipeline adopted to answer RQ.4. Steps annotated in black (e.g.,~\bcircle{1}) are described in detail, while steps in gray (e.g.,~\gcircle{D}) are mentioned in Section \ref{sec:method-rqd}.}
    \label{fig:rq2-data-collection}
\end{figure*}

In this question, we investigate the potential impact that \lbc{s} have on client applications.
For this, we implemented a data-collection pipeline that obtains a list of real-world client applications using either \scikit, \numpy, or \pandas, extracts the method calls performed in these clients, and selects the call vulnerables to the reported \bc{s}.
Figure \ref{fig:rq2-data-collection} depicts this procedure; more details about each step are described in the remainder of this section.

\elsparagraph{\small \bcircle{1}} \textit{Clients Dataset:}
We studied some datasets and data collection strategies to identify the best fit for our needs~\cite{Pimentel2019, Wang2020a, Quaranta2021, Zhang2021, Grotov2022}.
We consider the dataset's size, available documentation, and the effort to replicate and adapt it to our context.
We selected Grotov et al.'s dataset~\cite{Grotov2022} due to two reasons: (a) it comes in a structured format that can be queried using SQL; and (b) the authors also provide a tool that analyzes the source code of Jupyter Notebook and Python scripts, called Matroskin.
This dataset contains 847,881 preprocessed Jupyter Notebooks written in Python, extracted from GitHub between September and October 2020.

\elsparagraph{\small \bcircle{2}} \textit{Filtering Library Clients:}
In this step we selected all notebooks importing any of the libraries chosen from the initially obtained dataset.
We relied on the metadata available in the preprocessed database, which contains the list of imported modules for each client notebook.
We queried the database for all notebooks containing \textit{``sklearn''}, \textit{``numpy''}, or \textit{``pandas''} strings in their import list.
We detected 194,099 notebooks importing \scikit, 584,995 importing \numpy, and 348,899 importing \pandas~(\gcircle{D}).

\elsparagraph{\small \bcircle{3}} \textit{Calls Extraction:}
In this step, we extracted all existing method calls performed in each notebook.
However---unlike the list of imports---this information is not available by default in the dataset.
Instead, the authors provide only the number of method calls that belong to external sources, i.e.,~imported and third-party modules.
To obtain the actual method calls, we instrumented \textit{Matroskin} (\gcircle{M}) to extract all external calls during its syntactical analysis and re-executed it on the notebooks of \textit{Clients Dataset} (\gcircle{D}).
We ended up with 17,436,073 \textit{external calls} extracted from \scikit's clients (\gcircle{C}).
For \numpy\ and \pandas, we detected 43,016,995 and 24,156,627 calls, respectively.

\elsparagraph{\small \bcircle{4}} \textit{\bc\ Matching:}
In the last step of this pipeline, we selected all method calls vulnerable to \bc{s}.
Traditionally this could be achieved by tracking down the declaration of the called method, retrieving its arguments, and checking if the default argument is assigned in the call.
However, it is not straightforward to infer this information since Python is a dynamically-typed language~\cite{Tan2020, Grotov2022, Haryono2021}.
Due to this reason, we worked on a static matching heuristic to detect calls to methods identified as \bc{}.

\begin{figure*}[t!]
    \centering
    \includegraphics[width=1\linewidth]{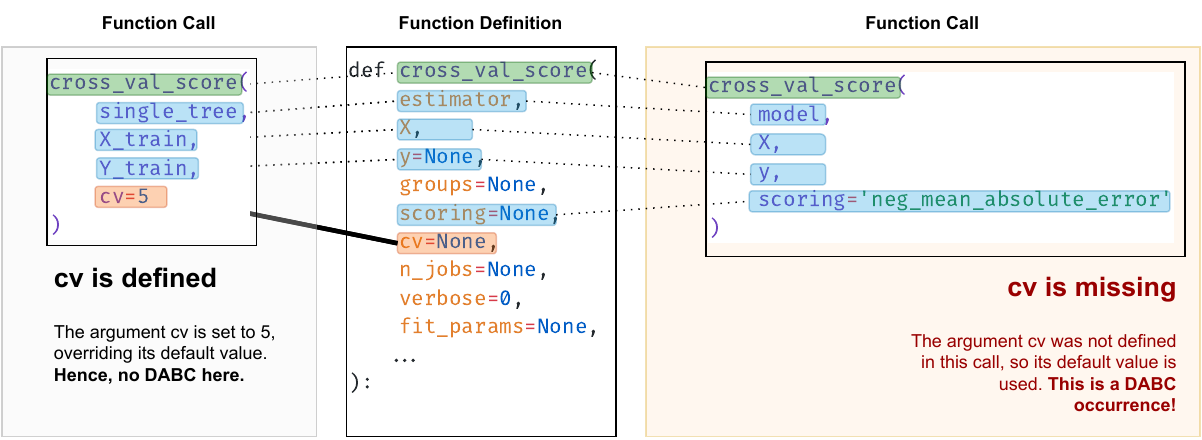}
    \caption{Illustrative example of how \bc{} Matching algorithm works.}
    \label{fig:bc-matching-algorithm}
\end{figure*}

The heuristic works as follows.
We first parse\footnote{We used the Python \mintinline{python}|gast| module at (\url{https://pypi.org/project/gast/}).} the definition of all 93 \textit{\bc{s}} (\gcircle{D}) identified in Section \ref{sec:method-rqa} and extract their class name (if any), method name, and list of defined arguments.
Next, for each \textit{external call} (\gcircle{C}), we parse and extract its method name and list of argument values; note that we can not directly obtain class names as Python is dynamically-typed.
Then we match \bc{s} and calls based on two conditions: (a) the \bc{} class name---if it exists---is in the same file where the call was retrieved; and (b) the method name in both \bc{} and in the call are the same.
For each successful match, we pair the call's argument values to the \bc{}'s defined arguments by assigning all positional arguments in sequence and assigning all keywords arguments based on the key provided.
Lastly, we check if the \bc{}'s default argument is assigned.
The call is considered vulnerable if \textbf{no value is assigned to the \bc\ argument}.
In practice, the call did not provide a value for it, so it relies on the \bc{}'s default argument value.
Figure \ref{fig:bc-matching-algorithm} exemplifies this heuristic using the \bc\ contained in \mintinline{python}|cross_val_score()| function, which changed \mintinline{python}|cv| default argument \textit{``from 3-fold to 5-fold''}.

\elsparagraph{\small \bcircle{5}} \textit{\bc{s} Calls Dataset:}
The \bc{} matching procedure identified 317,648 vulnerable calls for \scikit, the first library analyzed.
For \numpy\ and \pandas, we detected 1,275 and 172,152 calls, respectively.
We test the effectiveness of this heuristic by manually analyzing a randomly selected sample of 384 calls, equally divided among three authors.\footnote{The sample size was determined considering 95\% confidence level and 5\% confidence interval.}
They verified whether both method's call and \bc{} point to the same function and if the call is vulnerable.
To ensure the authors followed a similar verification pattern, they analyzed 38 calls together (10\% of the sample size).
In their evaluations, the authors identified 366 calls ($95.3\%$, $\pm 5\%$) as valid ones.
From the remaining 18, the heuristic mostly fail at detecting arguments outside the function call; e.g.,~arguments that were passed inside a Python dictionary, instead.
Such issue is beyond identification in static analysis, hence out of scope in our heuristic.

\section{Results}
\label{sec:results}

We present in this section the results for RQ.1--RQ.4 for \scikit, \numpy, and \pandas.
For each RQ, we first present the results for \scikit---it was the first library analyzed~\cite{Montandon2023}---and compare them with the results obtained for \numpy\ and \pandas\ in our extended work.

\subsection{\rqA}
\label{sec:results-rqa}

\paragraph{\scikit}
The 77 \bc{s} are spread over 61 distinct methods.
From these, 19 methods are declared outside of any class---e.g.,~\mintinline{python}|cross_validate()|, \mintinline{python}|k_means()|, etc---and account for 21 \bc{s}.
All 42 remaining methods declared in classes are constructors; they are responsible for 56 \bc{s}.
This finding emphasizes the central role that method constructors play when configuring \scikit\ models.

\begin{table}[ht!]
    \small
    \centering
    \caption{Top 10 most referred arguments in \scikit.}
    \label{tab:rq1-scikit}
    \begin{tabular}{lrr}
        \toprule
        \multirow{2}{*}{\textbf{Changed Argument}} & \multicolumn{2}{r}{\textbf{DABCs}}                 \\ \cline{2-3}
                                                   & \textbf{\# }                       & \textbf{\% }  \\
        \midrule
        \mintinline{python}|cv|                    & 20                                 & 26.0          \\
        \mintinline{python}|n_jobs|                & 8                                  & 10.4          \\
        \mintinline{python}|max_features|          & 6                                  & 7.8           \\
        \mintinline{python}|gamma|                 & 5                                  & 6.5           \\
        \mintinline{python}|n_estimators|          & 5                                  & 6.5           \\
        \mintinline{python}|n_splits|              & 4                                  & 5.2           \\
        \mintinline{python}|init|                  & 4                                  & 5.2           \\
        \mintinline{python}|multi_class|           & 3                                  & 3.9           \\
        \mintinline{python}|return_train_score|    & 3                                  & 3.9           \\
        \mintinline{python}|algorithm|             & 2                                  & 2.6           \\
        \midrule
        \textbf{Total}                             & \textbf{60}                        & \textbf{77.9} \\
        \bottomrule
    \end{tabular}
\end{table}

Individually, the methods \mintinline{python}|GridSearchCV.__init__()| and \mintinline[breaklines, breakafter=a]{python}|RandomizedSearchCV.__init__()| lead the rank of \bc{s} containing three occurrences, each.
Both \mintinline{python}|GridSearchCV| and \mintinline{python}|RandomizedSearchCV| classes implement strategies for optimizing ML models.
Moreover, both methods are vulnerable to changes performed in the same arguments: (i) \mintinline{python}|cv| defines the cross-validation strategy; (ii) \mintinline{python}|n_jobs| determines the number of jobs to run in parallel; and (iii) \mintinline{python}|return_train_score| determines if the method returns the computed training scores.
Twelve other methods have two \bc{s} each.
The remaining 47 methods show up with one \bc\ only.

We also analyzed the distribution of \bc{s} among the changed arguments.
In total, 24 arguments had their default value changed by at least one \bc{}.
Table \ref{tab:rq1-scikit} lists the top 10 most modified ones.
The \mintinline{python}|cv| argument stands out with 20 occurrences (26\%).
This argument defines the cross-validation strategy to split the data during model training and validation.
It is widely used in \scikit, as most supervised models rely on data-splitting techniques when they are trained.
Similarly, \mintinline{python}|n_jobs| (8 occurrences, 10.4\%) is also adopted in different scenarios.
The remaining arguments belong to specific classes and models.
For instance, \mintinline{python}|max_features| (6 occurrences, 7.8\%) and \mintinline{python}|n_estimators| (5 occurrences, 6.5\%) configure tree-based models.

\begin{table}[t]
    \caption{Most referred arguments in \numpy\ and \pandas.}
    \label{tab:rq1-numpy-pandas}
    \begin{subtable}[h]{0.45\textwidth}
        \small
        \centering
        \caption{\numpy}
        \label{tab:rq1-numpy}
        \begin{tabular}{lrr}
            \toprule
            \multirow{2}{*}{\textbf{Changed Argument}} & \multicolumn{2}{r}{\textbf{DABCs}}                \\ \cline{2-3}
                                                       & \textbf{\# }                       & \textbf{\% } \\
            \midrule
            \mintinline{python}|allow_pickle|          & 3                                  & 60.0         \\
            \mintinline{python}|axis|                  & 1                                  & 20.0         \\
            \mintinline{python}|rcond|                 & 1                                  & 20.0         \\
            \midrule
            \textbf{Total}                             & \textbf{5}                         & \textbf{100} \\
            \bottomrule
        \end{tabular}
    \end{subtable}
    \hfill
    \begin{subtable}[h]{0.45\textwidth}
        \small
        \centering
        \caption{\pandas.}
        \label{tab:rq1-pandas}
        \begin{tabular}{lrr}
            \toprule
            \multirow{2}{*}{\textbf{Changed Argument}} & \multicolumn{2}{r}{\textbf{DABCs}}                \\ \cline{2-3}
                                                       & \textbf{\# }                       & \textbf{\% } \\
            \midrule
            \mintinline{python}|auth_local_webserver|  & 2                                  & 18.2         \\
            \mintinline{python}|sort|                  & 2                                  & 18.2         \\
            \mintinline{python}|convert_axes|          & 1                                  & 9.1          \\
            \mintinline{python}|keep_tz|               & 1                                  & 9.1          \\
            \mintinline{python}|align|                 & 1                                  & 9.1          \\
            \mintinline{python}|hrules|                & 1                                  & 9.1          \\
            \mintinline{python}|inclusive|             & 1                                  & 9.1          \\
            \mintinline{python}|cache|                 & 1                                  & 9.1          \\
            \mintinline{python}|limit_direction|       & 1                                  & 9.1          \\
            \midrule
            \textbf{Total}                             & \textbf{11}                        & \textbf{100} \\
            \bottomrule
        \end{tabular}
    \end{subtable}
\end{table}

\paragraph{\numpy\ and \pandas}
Table \ref{tab:rq1-numpy-pandas} lists the most common arguments found for \numpy\ and \pandas.
Three \bc{s} (60\%) share the same argument modification in \numpy: \mintinline{python}|allow_pickle|.
This argument is used to enable Python pickle module, and was modified in response to a security issue found in this component.\footnote{\url{https://nvd.nist.gov/vuln/detail/CVE-2019-6446}}
In \pandas, the 11 \bc{s} are distributed among nine distinct arguments.
Just two arguments have been modified more than once: \mintinline{python}|auth_local_webserver| and \mintinline{python}|sort|, with two occurrences each.
The remainder seven arguments were detected in one \bc\ only.
Only one \bc\ was found in class constructors for \numpy\ and \pandas; a contrast with \scikit.
But most \bc{s} belong to methods declared in classes; three in \numpy\ (60\%) and seven in \pandas\ (63\%).

\begin{formal}
    While the 77 \bc{s} are spread across 24 arguments in \scikit, three and nine arguments concentrate all \bc{s} in \numpy\ and \pandas, respectively.
    Interestingly, \scikit\ reported most \bc{s} in class constructors, while \numpy\ and \pandas\ have only one.
\end{formal}

\subsection{\rqB}
\label{sec:results-rqb}

\begin{table}[ht!]
    \small
    \centering
    \caption{\bc{s} for each version for \scikit, \numpy, and \pandas.}
    \label{tab:versions}
    \begin{subtable}{1\linewidth}
        \centering
        \caption{\scikit}
        \label{tab:versions-scikit}
        \begin{tabular}{llrr}
            \toprule
            \multirow{2}{*}{\textbf{Type}} & \multirow{2}{*}{\textbf{Ver.}} & \multicolumn{2}{r}{\textbf{DABCs}}                \\ \cline{3-4}
                                           &                                & \textbf{\# }                       & \textbf{\% } \\ \midrule
            \textit{major}                 & 0.19                           & 3                                  & 3.9          \\
            \textit{major}                 & 0.20                           & 11                                 & 14.3         \\
            \textit{major}                 & 0.21                           & 3                                  & 3.9          \\
            \textit{major}                 & 0.22                           & 43                                 & 55.8         \\
            \textit{major}                 & 0.23                           & 3                                  & 3.9          \\
            \textit{major}                 & 0.24                           & 2                                  & 2.6          \\
            \textit{major}                 & 1.0                            & 1                                  & 1.3          \\
            \textit{major}                 & 1.1                            & 11                                 & 14.3         \\ \midrule
            \textbf{Total}                 &                                & \textbf{77}                        & \textbf{100} \\ \bottomrule
        \end{tabular}
    \end{subtable}
    \hfill
    \begin{subtable}{0.45\linewidth}
        \centering
        \caption{\numpy}
        \label{tab:versions-numpy}
        \begin{tabular}{llrr}
            \toprule
            \multirow{2}{*}{\textbf{Type}} & \multirow{2}{*}{\textbf{Ver.}} & \multicolumn{2}{r}{\textbf{DABCs}}                \\ \cline{3-4}
                                           &                                & \textbf{\# }                       & \textbf{\% } \\ \midrule
            \textit{minor}                 & 1.13.0                         & 1                                  & 20.0         \\
            \textit{minor}                 & 1.14.0                         & 1                                  & 20.0         \\
            \textit{bugfix}                & 1.16.3                         & 3                                  & 60.0         \\
            \midrule
            \textbf{Total}                 &                                & \textbf{5}                         & \textbf{100} \\ \bottomrule
        \end{tabular}
    \end{subtable}
    \hfill
    \begin{subtable}{0.45\linewidth}
        \centering
        \caption{\pandas}
        \label{tab:versions-pandas}
        \begin{tabular}{llrr}
            \toprule
            \multirow{2}{*}{\textbf{Type}} & \multirow{2}{*}{\textbf{Ver.}} & \multicolumn{2}{r}{\textbf{DABCs}}                \\ \cline{3-4}
                                           &                                & \textbf{\# }                       & \textbf{\% } \\ \midrule
            ---                            & 0.25.0                         & 2                                  & 18.2         \\
            \textit{major}                 & 1.0.0                          & 3                                  & 27.2         \\
            \textit{minor}                 & 1.1.0                          & 1                                  & 9.1          \\
            \textit{minor}                 & 1.3.0                          & 1                                  & 9.1          \\
            \textit{minor}                 & 1.4.0                          & 2                                  & 18.2         \\
            \textit{minor}                 & 1.5.0                          & 2                                  & 18.2         \\
            \midrule
            \textbf{Total}                 &                                & \textbf{11}                        & \textbf{100} \\ \bottomrule
        \end{tabular}
    \end{subtable}
\end{table}

Table \ref{tab:versions-scikit} presents the distribution of \bc{s} among each \scikit{}'s release.
They are distributed in eight versions;
the first ones appeared on version 0.19, released in November 2017.
Since then, we have identified \bc{s} in all major releases, i.e.,~no \bc\ was reported in minor versions.

Three versions concentrate 65 occurrences, representing 84.4\% of all \bc{s} reported in this study.
Specifically, version 0.22 stands out with 43 modifications in default arguments (55.8\%); both versions 0.20 and 1.1 appear next with 11 (14.3\%) \bc{s}.
The five remaining versions gather 12 \bc{s} in total.

We inspected in detail versions 0.22, 0.20, and 1.1 to better understand the reason for such disparity.
We find that the changes in these versions deal with \scikit{}'s popular features.
For instance, 19 out of 43 occurrences in 0.22 deal with \mintinline{python}|cv|.
Other five occurrences modify \mintinline{python}|n_estimators| argument.
Version 0.20 presents a similar characteristic, as eight occurrences point to the \mintinline{python}|n_jobs| argument.
Differently, \bc{s} are regularly distributed in version 1.1 with four distinct changes varying between two and three occurrences.

\paragraph{\numpy\ and \pandas}
The \numpy\ library has five \bc{s} distributed among three versions.
From these, three occurrences (60\%) are in version 1.16.3, a \textit{bugfix} release.
These are exactly the \bc{s} responsible for changing the \mintinline{python}|allow_pickle| argument to fix the security issue.

\pandas\ has 11 \bc{s} distributed among six versions.
As discussed in Section \ref{sec:method-rqb}, the library started adopting semantical versioning policies from version 1.0.0 onwards.
In this context, we observe that the \bc{s} were introduced in six subsequent minor versions; a noteworthy contrast when compared with \scikit\ and the literature.
Indeed, \pandas\ maintainers acknowledge they \textit{``will sometimes make behavior changing bug fixes, as part of minor or patch releases. Whether or not a change is a bug fix or an API-breaking change is a judgement call.''}\footnote{\url{https://pandas.pydata.org/docs/development/policies.html\#policies-version}}

\begin{formal}
    \scikit\ has \bc{s} in major versions, only.
    Differently, \numpy\ and \pandas\ have \bc{s} in minor versions.
    The reason for introducing \bc{s} in minor versions varies between the libraries.
    \numpy\ developers introduced them to fix security issues, while \pandas\ maintainers are more flexible in this regard.
\end{formal}

\subsection{\rqC}
\label{sec:results-rqc}

\begin{table}[ht!]
    \small
    \centering
    \caption{\bc{s} by module for \scikit, \numpy, and \pandas.}
    \label{tab:modules}
    \begin{subtable}{1\linewidth}
        \centering
        \caption{\scikit}
        \label{tab:modules-scikit}
        \begin{tabular}{lrr}
            \toprule
            \multirow{2}{*}{Modules} & \multicolumn{2}{c}{\textbf{DABCs}}                \\ \cline{2-3}
                                     & \textbf{\# }                       & \textbf{\% } \\ \midrule
            Dataset                  & 1                                  & 1.3          \\
            Data preprocessing       & 2                                  & 2.6          \\
            Data Decomposition       & 5                                  & 6.5          \\
            Data Analysis            & 2                                  & 2.6          \\
            Feature Processing       & 1                                  & 1.3          \\
            Model Training           & {42}                               & {54.5}       \\
            Model Evaluation         & {19}                               & {24.7}       \\
            Utils                    & 1                                  & 1.3          \\
            Pipeline                 & 3                                  & 3.9          \\
            Others                   & 1                                  & 1.3          \\
            \midrule
            \textbf{Total}           & \textbf{77}                        & \textbf{100} \\
            \bottomrule
        \end{tabular}
    \end{subtable}
    \\[0.25cm]
    \begin{subtable}{0.45\linewidth}
        \centering
        \caption{\numpy}
        \label{tab:modules-numpy}
        \begin{tabular}{lrr}
            \toprule
            \multirow{2}{*}{Modules} & \multicolumn{2}{c}{\textbf{DABCs}}                \\ \cline{2-3}
                                     & \textbf{\# }                       & \textbf{\% } \\ \midrule
            General Functions        & 3                                  & 60.0         \\
            Linear Algebra           & 1                                  & 20.0         \\
            Masked Arrays            & 1                                  & 20.0         \\
            \midrule
            \textbf{Total}           & \textbf{5}                         & \textbf{100} \\
            \bottomrule
        \end{tabular}
    \end{subtable}
    \hfill
    \begin{subtable}{0.45\linewidth}
        \centering
        \caption{\pandas}
        \label{tab:modules-pandas}
        \begin{tabular}{lrr}
            \toprule
            \multirow{2}{*}{Modules} & \multicolumn{2}{c}{\textbf{DABCs}}                \\ \cline{2-3}
                                     & \textbf{\# }                       & \textbf{\% } \\ \midrule
            DataFrame                & 3                                  & 27.2         \\
            General Functions        & 2                                  & 18.2         \\
            Input/Output             & 2                                  & 18.2         \\
            Style                    & 2                                  & 18.2         \\
            Index Objects            & 1                                  & 9.1          \\
            Series                   & 1                                  & 9.1          \\
            \midrule
            \textbf{Total}           & \textbf{11}                        & \textbf{100} \\
            \bottomrule
        \end{tabular}
    \end{subtable}
\end{table}

Table~\ref{tab:modules} describes the classified modules in all three libraries.
For \scikit, we observe that both \textit{Model Training} and \textit{Model Evaluation} stand out from the other modules, with 42 and 19 \bc{s}, respectively; together, they represent more than 79\% of all identified \bc{s}.
We hypothesize this concentration happens due to two reasons.
First, these modules are at the core of the library, as they contain the machine learning models and algorithms provided by the library.
Second, these classes present a high number of arguments---\mintinline{python}|SVC.__init__()| has 14 arguments, as shown in Section \ref{sec:background}---which increases the chances of having a default argument changed.
The remaining modules contain five changes or fewer.

\paragraph{\numpy\ and \pandas}
For \pandas, we identified six modules affected by the \bc{s}.
\textit{DataFrame} stands out with three \bc{s} (27.2\%).
This module contains essential functionalities to the \pandas\ library, as it provides access to the \texttt{DataFrame} class, used to represent any tabular data in the library.
Most \bc{s} in \numpy\ are in a general purpose module, a contrast with \scikit\ and \pandas.
These are precisely the \bc{s} which fixed the python \mintinline{python}|pickle| module's security issue.

\begin{formal}
    Both \scikit\ and \pandas\ have bc{s} concentrated in modules essential to their users: \textit{Model Training} and \textit{Model Evaluation} for \scikit, and \textit{DataFrame} for \pandas.
    \numpy, on the other hand, has most \bc{s} in a general-purpose module.
\end{formal}

\subsection{\rqD}
\label{sec:results-rqd}


\begin{table}[t!]
    \small
    \centering
    \caption{Most frequent \bc{s} calls in client applications.}
    \label{tab:dabcs-frequent-calls}
    \begin{subtable}{1.0\columnwidth}
        \small
        \centering
        \caption{\scikit}
        \label{tab:popular-scikit}
        \begin{tabular}{lrr}
            \toprule
            \multirow{2}{*}{\textbf{Class.Method(\textit{Default Argument})}}                               & \multicolumn{2}{c}{\textbf{Calls}}                                   \\ \cline{2-3}
                                                                                                            & \multicolumn{1}{c}{\textbf{\#}}    & \multicolumn{1}{c}{\textbf{\%}} \\
            \midrule
            \mintinline[escapeinside=@@]{python}|LogisticRegression.__init__(@\textit{multi\_class}@)|      & 38,323                             & 12.1                            \\
            \mintinline[escapeinside=@@]{python}|LogisticRegression.__init__(@\textit{solver}@)|            & 31,290                             & 9.9                             \\
            \mintinline[escapeinside=@@]{python}|RandomForestClassifier.__init__(@\textit{max\_features}@)| & 30,874                             & 9.7                             \\
            \mintinline[escapeinside=@@]{python}|SVC.__init__(@\textit{decision\_funciton\_shape}@)|        & 29,904                             & 9.4                             \\
            \mintinline[escapeinside=@@]{python}|GridSearchCV.__init__(@\textit{return\_train\_score}@)|    & 24,930                             & 7.8                             \\
            \mintinline[escapeinside=@@]{python}|SVC.__init__(@\textit{gamma}@)|                            & 22,258                             & 7.0                             \\
            \mintinline[escapeinside=@@]{python}|KMeans.__init__(@\textit{algorithm}@)|                     & 22,063                             & 6.9                             \\
            \mintinline[escapeinside=@@]{python}|r2_score(@\textit{multioutput}@)|                          & 20,805                             & 6.5                             \\
            \mintinline[escapeinside=@@]{python}|GridSearchCV.__init__(@\textit{n\_jobs}@)|                 & 16,396                             & 5.2                             \\
            \mintinline[escapeinside=@@]{python}|RandomForestRegressor.__init__(@\textit{max\_features}@)|  & 14,970                             & 4.7                             \\
            \midrule
            \textbf{Total}                                                                                  & \textbf{251,813}                   & \textbf{79.2}                   \\
            \bottomrule
        \end{tabular}
    \end{subtable}
    \begin{subtable}{1.0\columnwidth}
        \vspace*{2em}
        \small
        \centering
        \caption{\pandas}
        \label{tab:popular-pandas}
        \begin{tabular}{lrr}
            \toprule
            \multirow{2}{*}{\textbf{Class.Method(\textit{Default Argument})}}                         & \multicolumn{2}{c}{\textbf{Calls}}                                   \\ \cline{2-3}
                                                                                                      & \multicolumn{1}{c}{\textbf{\#}}    & \multicolumn{1}{c}{\textbf{\%}} \\ \midrule
            \mintinline[escapeinside=@@]{python}|concat(@\textit{sort}@)|                             & 108,647                            & 63.1                            \\
            \mintinline[escapeinside=@@]{python}|to_datetime(@\textit{cache}@)|                       & 51,631                             & 30.0                            \\
            \mintinline[escapeinside=@@]{python}|read_json(@\textit{convert\_axes}@)|                 & 9,367                              & 5.4                             \\
            \mintinline[escapeinside=@@]{python}|read_gbq(@\textit{auth\_local\_webserver}@)|         & 904                                & 0.5                             \\
            \mintinline[escapeinside=@@]{python}|DataFrame.append(@\textit{sort}@)|                   & 673                                & 0.4                             \\
            \mintinline[escapeinside=@@]{python}|Series.between(@\textit{inclusive}@)|                & 558                                & 0.3                             \\
            \mintinline[escapeinside=@@]{python}|DataFrame.to_gbq(@\textit{auth\_local\_webserver}@)| & 297                                & 0.2                             \\
            \mintinline[escapeinside=@@]{python}|DatetimeIndex.to_series(@\textit{keep\_tz}@)|        & 47                                 & $> 0.1$                         \\
            \mintinline[escapeinside=@@]{python}|Styler.bar(@\textit{align}@)|                        & 28                                 & $> 0.1$                         \\
            \midrule
            \textbf{Total}                                                                            & \textbf{172,152}                   & \textbf{100}                    \\
            \bottomrule
        \end{tabular}
    \end{subtable}
    \begin{subtable}{1.0\columnwidth}
        \vspace*{2em}
        \small
        \centering
        \caption{\numpy}
        \label{tab:popular-numpy}
        \begin{tabular}{lrr}
            \toprule
            \multirow{2}{*}{\textbf{Class.Method(\textit{Default Argument})}}          & \multicolumn{2}{c}{\textbf{Calls}}                                   \\ \cline{2-3}
                                                                                       & \multicolumn{1}{c}{\textbf{\#}}    & \multicolumn{1}{c}{\textbf{\%}} \\ \midrule

            \mintinline[escapeinside=@@]{python}|lstsq(@\textit{rcond}@)|              & 1,270                              & 99.6                            \\
            \mintinline[escapeinside=@@]{python}|read_array(@\textit{allow\_pickle}@)| & 5                                  & 0.4                             \\
            \midrule
            \textbf{Total}                                                             & \textbf{1,275}                     & \textbf{100}                    \\
            \bottomrule
        \end{tabular}
    \end{subtable}
\end{table}

\subsubsection{What are the most frequent \bc{s}?}
Table \ref{tab:popular-scikit} lists the most frequent \bc{s} identified in \scikit\ client applications.
We detected vulnerable calls for 72 out of the 77 \bc{s} identified previously (93\%);
The top 10 gathered 251,813 of the 317,648 vulnerable calls (79.2\%), suggesting a heavy-tail distribution.
While the 10th most frequent \bc\ contains 14,970 vulnerable calls, the median value is 365.

We observe that nine vulnerable calls refer to class constructors.
This highlights a common practice in \scikit\ where most configuration arguments are passed when creating the model instance.
The first two calls in the table refer to the same method (\mintinline{python}|LogisticRegression.__init__()|), but with different argument values (\mintinline{python}|multi_class| and \mintinline{python}|solver|).
Same behavior applies for \mintinline{python}|SVC.__init__()| (\mintinline{python}|decision_function_shape| and \mintinline{python}|gamma|) and \mintinline{python}|GridSearchCV.__init__()| (\mintinline{python}|return_train_score| and \mintinline{python}|n_jobs|) methods.
We also observe the same default argument used in two distinct classes: \mintinline{python}|max_features| is used when calling \mintinline{python}|RandomForestClassifier.__init__()| and \mintinline[breaklines, breakafter=t]{python}|RandomForestRegressor.__init__()| methods.

\paragraph{\numpy\ and \pandas}
Tables \ref{tab:popular-numpy} and \ref{tab:popular-pandas} list the most frequent \bc{s} calls for both libraries.
We detected vulnerable calls in nine out of 11 \bc{s} for \pandas\ (81\%) and in two out of five \bc{s} for \numpy\ (40\%).
The concentration of vulnerable calls are even higher in these libraries.
While \mintinline{python}|concat(sort)| and \mintinline{python}|to_datetime(cache)| are responsible for 93\% of all occurrences in \pandas, \mintinline{python}|lstsq(rcond)| is solely responsible for 99.6\% of all vulnerable calls in \numpy.

\begin{formal}
    Few \bc{s} are responsible for most vulnerable calls in client applications, following a heavy-tail distribution.
    While 12\% (10 out of 77) of \scikit\ \bc{s} are responsible for 80\% of vulnerable calls, 18\% (2 out of 11) and 20\% (1 out of 5) of \numpy\ and \pandas\ \bc{s} concentrate 93\% and 99\%, respectively.
\end{formal}

\subsubsection{How many clients are vulnerable to \bc{s}?}

Table \ref{tab:replication-clients-numbers} summarizes the number of clients vulnerable in each library.
For \scikit, 67,747 out of 194,099 clients are vulnerable to at least one \bc\ (35\%).
Despite the higher number of clients in \pandas\ and \numpy, these numbers do not translate into more potential vulnerabilities.
We detected 73,469 vulnerable clients in \pandas\ (21\%) and only 738 in \numpy\ (0.13\%).
Their application context may influence these results.
Scikit-Learn implements complex functionalities to handle a wide range of scenarios, including training and feature engineering techniques, as well as machine learning algorithms.
These functionalities are highly sensitive to parameter calibration, which can lead to unpredictable outcomes.
On the other hand, Pandas and NumPy provide arguments that yield more predictable results.

\begin{table}[ht!]
    \small
    \centering
    \caption{Number of vulnerable clients and API calls from \pandas\ and \numpy\ in comparison with \scikit.}
    \label{tab:replication-clients-numbers}
    \begin{tabular}{lr|rr}
        \toprule
                                       & \textbf{Scikit-Learn} & \textbf{Pandas} & \textbf{NumPy} \\ \midrule
        \textbf{\# All Clients}        & 194,099               & 348,899         & 584,995        \\
        \textbf{\# Vulnerable Clients} & 67,747                & 73,469          & 738            \\
        \textbf{\# Vulnerable Calls}   & 317,648               & 172,152         & 1,275          \\ \bottomrule
    \end{tabular}
\end{table}

\paragraph{Correlation with Structural Metrics}

We triangulated the number of vulnerable calls with the 15 structural metrics collected by Grotov et al.~\cite{Grotov2022} to verify how traditional software metrics relate to \bc{s}.
This is a first step towards understanding how software quality influences the emergence of \bc{s}.
For this, we executed the Spearman correlation test between the number of calls and each metric separately.
We opted for Spearman due to its robustness in interpreting non-normalized distributions~\cite{Hinkle2003}.
Following the guidelines proposed in other works~\cite{Borges2018, Montandon2019b, Tan2020}, we interpret its coefficient according to the following: $0.00 \leq negligible < 0.30 \leq low < 0.50 \leq moderate < 0.70 \leq high < 0.90 \leq very high < 1.00$.

\begin{table}[ht!]
    \footnotesize
    \centering
    \caption{Spearman correlation between the structural metrics collected by Grotov et al.~\cite{Grotov2022} and the number of calls vulnerable to \bc{s} in client applications. The bullet shape quantifies the correlation level: negligible ($\bullet$), low ($\bullet\bullet$), and moderate ($\bullet\bullet\bullet$).}
    \label{tab:clients-correlated-metrics}
    \begin{tabular}{lrc}
        \toprule
        \multirow{2}{*}{\textbf{Metric}} & \multicolumn{2}{c}{\textbf{Correlation}}                           \\ \cline{2-3}
                                         & \multicolumn{1}{c}{\textbf{Coeff.}}      & \textbf{Level}          \\
        \midrule
        \textsc{Code Writing}            &                                          &                         \\[0.20cm]
        \textbf{SLOC}                    & \textbf{0.38}                            & $\bullet\bullet$        \\
        Blank LOC                        & 0.28                                     & $\bullet$               \\
        Extended comments LOC            & 0.24                                     & $\bullet$               \\
        Comments LOC                     & 0.19                                     & $\bullet$               \\
        \midrule
        \textsc{Function Usage}          &                                          &                         \\[0.20cm]
        \textbf{API functions (count)}   & \textbf{0.50}                            & $\bullet\bullet\bullet$ \\
        API functions (unique)           & 0.38                                     & $\bullet\bullet$        \\
        Other functions (count)          & 0.32                                     & $\bullet\bullet$        \\
        Built-in functions (count)       & 0.29                                     & $\bullet$               \\
        Built-in functions (unique)      & 0.19                                     & $\bullet$               \\
        User-defined functions (count)   & 0.18                                     & $\bullet$               \\
        User-defined functions (unique)  & 0.15                                     & $\bullet$               \\
        \midrule
        \textsc{Complexity}              &                                          &                         \\[0.20cm]
        \textbf{Cell coupling}           & \textbf{0.41}                            & $\bullet\bullet$        \\
        Function coupling                & 0.14                                     & $\bullet$               \\
        NPAVG                            & 0.13                                     & $\bullet$               \\
        Cyclomatic complexity            & 0.09                                     & $\bullet$               \\
        \bottomrule
    \end{tabular}
\end{table}

Table \ref{tab:clients-correlated-metrics} presents the correlation results for \scikit;
we marked in bold the top three metrics with higher correlation coefficients.
Overall, we did not identify any high correlation between the structural metrics and the number of vulnerable calls in client applications.
On the contrary, the correlation levels of most metrics are either $low$ (four) or $negligible$ (ten).
Only \textit{API functions (count)} presented a $moderate$ correlation with the number of vulnerable calls (0.50); the correlation with \textit{API functions (unique)} is lower, though (0.38, $low$ level).
\textit{Complexity}-based metrics are independent to the number of \bc{s} calls: \textit{Cyclomatic complexity}, \textit{Function coupling}, and \textit{NPAVG} scored the lowest correlation coefficients with 0.09, 0.13, and 0.14, respectively.
These findings suggest that the presence of \bc{s} is relatively independent of the codebase size, functions usage, and complexity.

The correlation coefficients for \numpy\ and \pandas\ are presented in Table \ref{tab:replication-correlated-metrics}, where we added a $\Delta$ column comparing their results with \scikit.
Overall, the correlation presented lower coefficients for both libraries.
\pandas\ reported just two metrics at \textit{low} level, while the remaining ones are considered \textit{negligible}.
For \numpy, we discarded the correlation results for seven metrics as they reported \textit{p-value} higher than 0.05 (marked as ``---''); the remaining ones are \textit{negligible}.
Interestingly, we observe a significant difference between the coefficients of \textit{API functions}-based metrics when compared with \scikit; around 30 points lower in \pandas, and no correlation confirmed for \numpy.

\begin{table*}[t!]
    \footnotesize
    \centering
    \caption{Spearman correlation between the structural metrics collected by Grotov et al.~\cite{Grotov2022} and the number of calls vulnerable to \bc{s} in client applications. The bullet shape quantifies the correlation level: negligible ($\bullet$), low ($\bullet\bullet$), and moderate ($\bullet\bullet\bullet$). $\Delta$ reports the difference with \scikit\ coefficients (Table \ref{tab:clients-correlated-metrics}).}
    \label{tab:replication-correlated-metrics}
    \begin{tabular}{l|rcc|rcc}
        \toprule
        \multirow{2}{*}{\textbf{Metric}} & \multicolumn{3}{c|}{\pandas}        & \multicolumn{3}{c}{\numpy}                                                                              \\
                                         & \multicolumn{1}{c}{\textbf{Coeff.}} & \textbf{Level}             & $\Delta$ & \multicolumn{1}{c}{\textbf{Coeff.}} & \textbf{Level} & $\Delta$ \\ \midrule
        \textsc{Code Writing}            &                                     &                            &          &                                     &                &          \\[0.20cm]
        SLOC                             & \textbf{0.31}                       & $\bullet\bullet$           & -0.09    & \textbf{0.17}                       & $\bullet$      & -0.21    \\
        Blank LOC                        & 0.21                                & $\bullet$                  & -0.09    & 0.15                                & $\bullet$      & -0.13    \\
        Extended comments LOC            & 0.19                                & $\bullet$                  & -0.06    & 0.13                                & $\bullet$      & -0.11    \\
        Comments LOC                     & 0.19                                & $\bullet$                  & 0.00     & 0.12                                & $\bullet$      & -0.07    \\ \midrule
        \textsc{Function Usage}          &                                     &                            &          &                                     &                &          \\[0.20cm]
        API functions (count)            & 0.12                                & $\bullet$                  & -0.32    & ---                                 & ---            & ---      \\
        API functions (unique)           & 0.08                                & $\bullet$                  & -0.30    & ---                                 & ---            & ---      \\
        Other functions (count)          & \textbf{0.34}                       & $\bullet\bullet$           & +0.02    & \textbf{0.23}                       & $\bullet$      & -0.09    \\
        Built-in functions (count)       & 0.21                                & $\bullet$                  & -0.08    & ---                                 & ---            & ---      \\
        Built-in functions (unique)      & 0.18                                & $\bullet$                  & -0.01    & 0.16                                & $\bullet$      & -0.03    \\
        User-defined functions (count)   & 0.12                                & $\bullet$                  & -0.06    & ---                                 & ---            & ---      \\
        User-defined functions (unique)  & 0.12                                & $\bullet$                  & -0.03    & ---                                 & ---            & ---      \\ \midrule
        \textsc{Complexity}              &                                     &                            &          &                                     &                &          \\[0.20cm]
        Cell coupling                    & \textbf{0.29}                       & $\bullet$                  & -0.12    & \textbf{0.24}                       & $\bullet$      & -0.17    \\
        Function coupling                & 0.15                                & $\bullet$                  & +0.01    & 0.09                                & $\bullet$      & -0.05    \\
        NPAVG                            & 0.03                                & $\bullet$                  & -0.10    & ---                                 & ---            & ---      \\
        Cyclomatic complexity            & 0.12                                & $\bullet$                  & +0.03    & ---                                 & ---            & ---      \\
        \bottomrule
    \end{tabular}
\end{table*}

\begin{formal}
    35\% of client applications are vulnerable to \bc{s} in \scikit, while 21\% and 0.13\% are vulnerable in \pandas\ and \numpy, respectively.
    Besides, we did not find any substantial correlation between source code metrics and \bc{s} calls for all three libraries.
\end{formal}

\subsubsection{Which versions make clients more vulnerable?}

\begin{figure}[ht!]
    \centering
    \includegraphics[width=0.7\columnwidth]{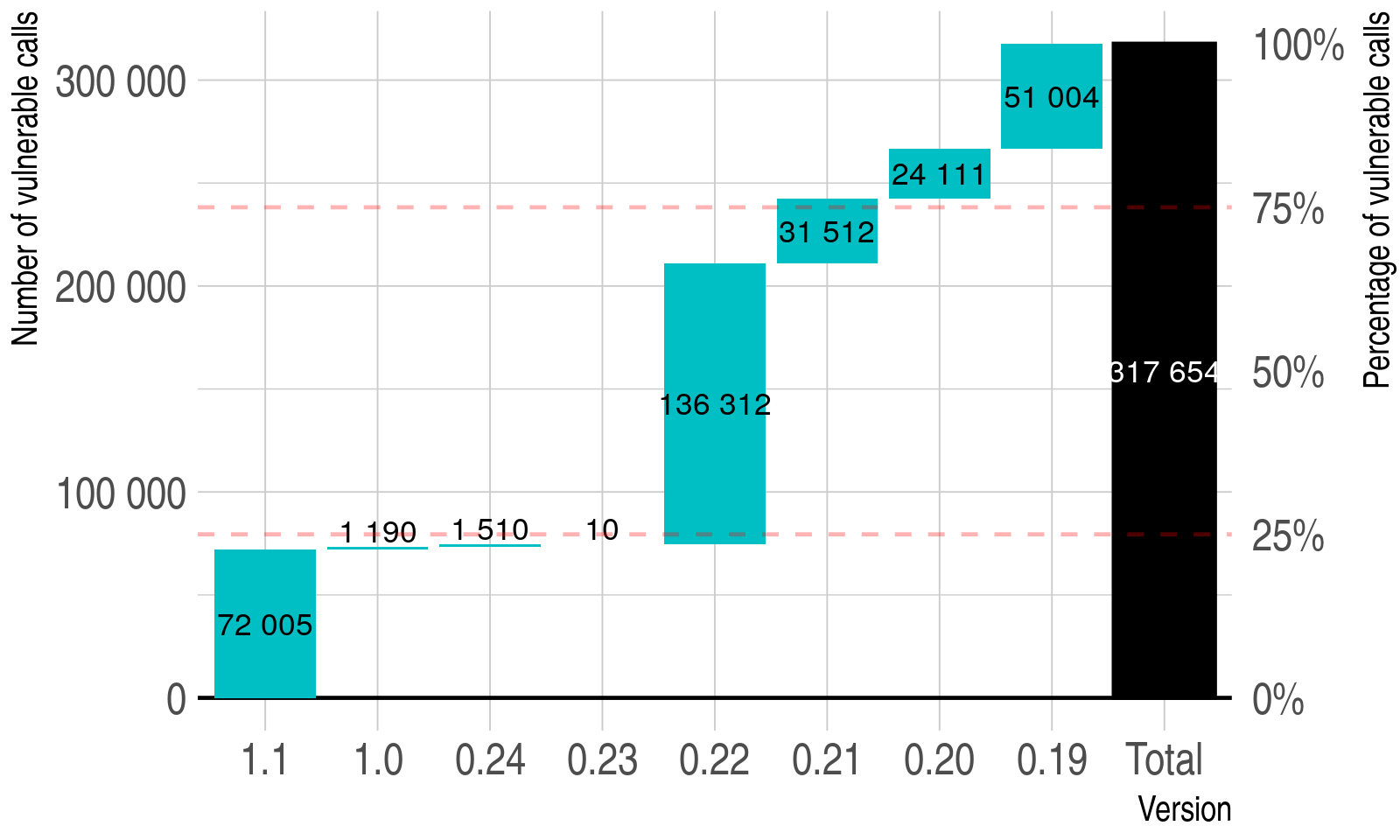}
    \caption{Number of vulnerable calls in each version in \scikit.}
    \label{fig:clients-calls-by-version}
\end{figure}

Figure \ref{fig:clients-calls-by-version} depicts the vulnerable calls in each \scikit\ version.
Versions 0.22, 1.1, and 0.19 stand out with 136,312 (42.9\%), 72,005 (22.7\%), and 51,004 (16.1\%) vulnerable calls, respectively; altogether, these versions concentrate 81.6\% of the vulnerable calls.
By contrast, versions 0.23, 1.0, and 0.24 had the most negligible impact on clients with 10 ($<$0.01\%), 1,190 (0.37\%), and 1,510 (0.48\%) vulnerable calls.

We observe that 23\% of vulnerable calls happen in more recent versions, i.e., version 1.0 onwards.
The proportion goes to 66\% when we extend this analysis to version 0.22.
In other words, two-thirds of all vulnerable calls are due to \bc{s} reported from versions 0.22 onwards.

\paragraph{\numpy\ and \pandas}
We observe that few versions bring together almost all calls we analyze.
For \pandas, versions 0.25.0 and 1.0.0 are responsible for 35.4\% and 63.5\% of vulnerable calls, respectively; version 1.14.0 is responsible for 99.6\% of them in \numpy.

\begin{formal}
    For \scikit, versions 0.22, 1.1, and 0.19 had the highest impact on clients, accounting for 81\% of vulnerable calls.
    Likewise, \pandas\ and \numpy\ gather most vulnerable calls in few versions: 0.25.0 and 1.0.0 account for 98.9\% of calls in \pandas, and 1.14.0 for 99.6\% in \numpy.
\end{formal}

\subsubsection{Which modules are more vulnerable in clients?}

Table \ref{tab:clients-modules} presents the number of vulnerable calls located in each \scikit\ module.
\textit{Model Training} and \textit{Model Evaluation} clearly stand out with 248,014 and 64,927 calls each; both modules condense 98.5\% of all vulnerable calls.
Such higher concentration indicates that most \bc{s} show up when dealing with the machine learning models.
On the other hand, no other module gathers more than 1\% of vulnerable calls; \textit{Pipeline} is the highest remaining one with 0.59\%.
We also verified the modules that are vulnerable together, i.e.,~in the same client.
In this perspective, 16,233 (24.0\%) clients are simultaneously vulnerable in two modules, 1,204 (1.8\%) in three, and 172 (0.2\%) in four modules.

\begin{table}[ht!]
    \small
    \centering
    \caption{Number of vulnerable calls for each \scikit\ module.}
    \label{tab:clients-modules}
    \begin{tabular}{lrr}
        \toprule
        \multirow{2}{*}{\textbf{Module}} & \multicolumn{2}{c}{\textbf{Calls}}                                   \\ \cline{2-3}
                                         & \multicolumn{1}{c}{\textbf{\#}}    & \multicolumn{1}{c}{\textbf{\%}} \\
        \midrule
        Data Analysis                    & 32                                 & 0.01                            \\
        Data Decomposition               & 1,067                              & 0.33                            \\
        Feature Processing               & 103                                & 0.03                            \\
        Model Evaluation                 & 64,927                             & 20.43                           \\
        Model Training                   & 248,014                            & 78.07                           \\
        Utils                            & 173                                & 0.05                            \\
        Dataset                          & 1,070                              & 0.33                            \\
        Pipeline                         & 1,876                              & 0.59                            \\
        Preprocessing                    & 392                                & 0.12                            \\
        \bottomrule
    \end{tabular}
\end{table}

\paragraph{\numpy\ and \pandas}
Table \ref{tab:replication-clients-modules} presents the number of vulnerable calls located in each \numpy\ and \pandas\ module.
\textit{General Functions} gathers 93.1\% of all vulnerable calls in \pandas, although it is responsible for two \bc{s}, only.
Besides containing core functions, the \textit{DataFrame} module is responsible for less than 1.0\% of all vulnerable calls.
As for \numpy, the \textit{Linear Algebra} module accounted for 99.6\% of all vulnerable calls in \numpy.

\begin{table}[ht!]
    \small
    \centering
    \caption{Vulnerable calls on \numpy\ and \pandas\ modules.}
    \label{tab:replication-clients-modules}
    \begin{subtable}{0.45\linewidth}
        \centering
        \caption{\numpy.}
        \begin{tabular}{lrr}
            \toprule
            \multirow{2}{*}{\textbf{Module}} & \multicolumn{2}{c}{\textbf{Calls}}                                   \\ \cline{2-3}
                                             & \multicolumn{1}{c}{\textbf{\#}}    & \multicolumn{1}{c}{\textbf{\%}} \\
            \midrule
            Linear Algebra                   & 1,270                              & 99.6                            \\
            General Functions                & 5                                  & 0.4                             \\
            Masked Arrays                    & 0                                  & 0.0                             \\
            \bottomrule
        \end{tabular}
    \end{subtable}
    \hfill
    \begin{subtable}{0.5\linewidth}
        \centering
        \caption{\pandas.}
        \begin{tabular}{lrr}
            \toprule
            \multirow{2}{*}{\textbf{Module}} & \multicolumn{2}{c}{\textbf{Calls}}                                   \\ \cline{2-3}
                                             & \multicolumn{1}{c}{\textbf{\#}}    & \multicolumn{1}{c}{\textbf{\%}} \\
            \midrule
            General Functions                & 160,278                            & 93.1                            \\
            Input/Output                     & 10,271                             & 5.8                             \\
            DataFrame                        & 970                                & 0.7                             \\
            Series                           & 558                                & 0.3                             \\
            Index Objects                    & 47                                 & $<$ 0.1                         \\
            Style                            & 28                                 & $<$ 0.1                         \\
            \bottomrule
        \end{tabular}
    \end{subtable}
\end{table}

\begin{formal}
    Few modules are responsible for most vulnerable calls in client applications.
    \textit{Model Training} and \textit{Model Evaluation} are responsible for 98.5\% of vulnerable calls in \scikit.
    In \pandas, \textit{General Functions} accounts for 93.1\% of all vulnerable calls, while \textit{Linear Algebra} is responsible for 99.6\% in \numpy.
\end{formal}

\section{Extended Analysis}
\label{sec:extended-analysis}

So far we describe \textit{what} \lbc{s} are, \textit{when} and \textit{where} they are introduced, as well as \textit{how} clients are vulnerable to them.
However, understanding \textit{why} maintainers perform such modifications remains open.
We claim that investigating this aspect is paramount as it could reveal typical scenarios where these values are modified.
Therefore, we extend our work to investigate the reasons that led API maintainers to perform these changes and, consequently, introduce \bc{s} to their APIs.
We define another two research questions focused on tackling this matter.
The remaining of this section describes the methodology and results we obtained for each question.

\subsection{\rqE}

\subsubsection{Methodology}

The goal of this question is to better understand the root causes that led maintainers to implement the changes that resulted in \bc{s}.
As we are dealing with open-source projects, we assume the reasons for these changes are documented in the projects' issue-tracking systems.
Hence, we designed a three-step procedure to analyze the issues and pull requests responsible for introducing the \bc{s}, described as follows.

\paragraph{Issue Mapping}
We rely on the \textit{dabc\_url} attribute (see Section \ref{sec:method-collection}) to map each \bc\ with the respective issue that introduced it.
Specifically, one author accessed each \textit{dabc\_url} individually and asked GitHub to blame the exact line that reported the \bc.
Next, the author read the commit messages from each blame, looking for numbers that indicate issues code, e.g., \textit{``DEPR: change pd.concat sort=None to sort=False (\texttt{\#29786})''}.
As all three libraries adopt a convention to include the issue number in their commit messages, we were able to map an issue number to all \textbf{93 \bc{s}}.

\paragraph{Comments Analysis}
Two authors examined the comments of each issue to detect the major reason that prompted maintainers to introduce the \bc.
Initially, they read the original issue thread looking for a comment that helps articulate the following sentence: \textit{``The \bc\ was introduced because \dots''}.
In our view, such a comment would contain the reason that led to the \bc{}'s introduction, hence answering our question.
If the comment is not found, the authors recursively apply the same procedure for issues linked in the thread that directly mentions the change performed in the \bc;
they stop looking as soon as they find a comment or when all related issues have been analyzed.
The authors identified comments justifying the \bc{} modification for \textbf{76 of them (81.7\%)}; 62 for \scikit, nine for \pandas, and five for \numpy.

\paragraph{Comments Classification}
Finally, we classified the issue comments into four high-level reasons, obtained from Brito et al.~\cite{Brito2020}.
We relied on these categories since the authors identified them after asking Java Developers why they introduced breaking changes to their APIs;
the same question analyzed in this RQ.
The categories are summarized as follows.
\begin{itemize}
    \item \textit{New Feature:} introduced to support new features implemented by the maintainers.
    \item \textit{API Compatibility:} aimed at simplifying the use of the library's API.
    \item \textit{Maintainability:} added with the goal of improving the API maintainability.
    \item \textit{Bug Fixing:} performed to fix a bug raised due to the old default value.
\end{itemize}

\subsubsection{Results}



Table \ref{tab:dabc-reasons} describes the number of \bc{s} classified in each category, for each library.
In absolute numbers, \textit{Maintainability}-based changes are the most frequent, responsible for introducing 54 \bc{s}.
\textit{Bug Fixing} and \textit{API Compatibility} appears next with 10 and nine \bc{s}, respectively.
By contrast, no reason was predominant across the libraries.
While \textit{Maintainability} is the reason for adding most \bc{s} in \scikit\ (80.7\%), \textit{API Compatibility} and \textit{Bug Fixing} are the prevalent ones for \pandas\ (44.4\%) and \numpy\ (60\%), respectively.
Only \textit{New Feature}-based modifications remained irrelevant in both cases; only one occurrence was reported for \scikit\ and \pandas.

\begin{table}[ht!]
    \centering
    \caption{Number of \bc{s} grouped by the reason they were introduced. Percentage values are calculated by each library.}
    \label{tab:dabc-reasons}
    \begin{tabular}{lccc}
        \toprule
        \textbf{Category}          & \scikit      & \pandas     & \numpy   \\
        \midrule
        \textit{New Feature}       & 1   (1.6\%)  & 1  (11.1\%) & --       \\
        \textit{API Compatibility} & 5   (8.1\%)  & 4  (44.4\%) & 1 (20\%) \\
        \textit{Maintainability}   & 50  (80.7\%) & 3  (33.3\%) & 1 (20\%) \\
        \textit{Bug Fixing}        & 6   (9.6\%)  & 1  (11.1\%) & 3 (60\%) \\
        \bottomrule
    \end{tabular}
\end{table}

\subsection{\rqF}

\subsubsection{Methodology}

This question aims at studying the effects these changes have on client applications.
We understand answering this question is important as \bc{s} affect client applications in distinct ways.
While the minimum working example presented in Figure \ref{lst:dabc-example} drastically modifies the client's outcome, other changes might not be so impactful.
For example, one \pandas's \bc{} just enables a \textit{``cache of unique, converted dates to apply the datetime conversion''.}\footnote{\url{https://pandas.pydata.org/pandas-docs/version/1.0/reference/api/pandas.to_datetime.html}.}

By relying on an open card sorting procedure~\cite{Spencer2009}, two authors read the description of all \bc{s}'s arguments and classified them according to the effect they produce when their function is executed.
For this, they analyzed and classified \pandas\ and \numpy\ \bc{s} to leverage an initial set of categories.
Next, they classified all \scikit\ \bc{s} using the previous ones whenever possible.
Finally, a third author independently classified the \bc{s} using the same categories.
The inter-rater agreement---calculated using Cohen's Kappa~\cite{Warrens2015}---was 0.76 on average, indicating a substantial agreement between the authors.
They discussed the conflicting cases to reach a common ground.
This procedure resulted in four categories, described below:

\begin{itemize}
    \item \textit{Aesthetics} refers to arguments affecting the format or style of the data returned by the function, i.e.,~do not change the structure or content of the data processed by the function.
          For example, the \mintinline{python}|align| argument of \pandas's \bc\ \#07 affects the alignment of the bars processed by the \mintinline{python}|Styler.bar()| function.\footnote{Described at \url{https://pandas.pydata.org/pandas-docs/version/2.2/reference/api/pandas.io.formats.style.Styler.bar.html}, argument \mintinline{python}|align|.}
          While its previous default value (\mintinline{python}|'left'|) drew the bars rightwards, the new defalut (\mintinline{python}|'mid'|) draws the bars at the center. Note that this modification just changes the appearance of the bars, not their structure or content.
    \item \textit{Behavior} refers to arguments whose changes led to different outcomes.
          This is the case involving the many \scikit\ \mintinline{python}|cv|'s \bc{s}, where its default value changed from \mintinline{python}|3| to \mintinline{python}|5|.\footnote{Described at \url{https://scikit-learn.org/stable/modules/generated/sklearn.model_selection.cross_val_score.html}, argument \mintinline{python}|cv|.}
          As expected, changing the value of \textit{cross-validation} necessarily impacts the performance of ML models.
    \item \textit{Performance} includes arguments responsible for customizing a function's performance.
          For instance, the argument \mintinline{python}|n_jobs| reported in some \scikit's \bc{s} allows clients to define the number of processing jobs to use in the computation.\footnote{Described at \url{https://scikit-learn.org/stable/modules/generated/sklearn.model_selection.GridSearchCV.htm}, argument \mintinline{python}|n_jobs|.} Its old default value was \mintinline{python}|1| and indicated that all computations would run in a single job by default. Later maintainers changed it to \mintinline{python}|None|, allowing these processes to use more jobs by default.
    \item \textit{Refactoring} involves \bc{s} arguments introduced as a consequence of a refactoring action.
          The \bc{s} based on the \mintinline{python}|max_features| argument illustrate this case.\footnote{Described at \url{https://scikit-learn.org/stable/modules/generated/sklearn.ensemble.RandomForestClassifier.html}, argument \mintinline{python}|max_features|.}
          Previously, all tree models relying on this argument had \mintinline{python}|'auto'| as their default value.
          In this scenario, the actual value of \mintinline{python}|max_features| would automatically be redefined based on the model's type; classification-based ones would use \mintinline{python}|'sqrt'|, and regression-based ones would use \mintinline{python}|1.0| as their default value.
          The \bc{s} resulted from this change replaced \mintinline{python}|'auto'| with the actual default according to each model; classification-based had its default changed to \mintinline{python}|'sqrt'|, and regression-based had its default changed to \mintinline{python}|1.0|.
          All-in-all, the default values for these classifications remained exactly the same, i.e.,~the change does not affect clients that use this function.
\end{itemize}

\subsubsection{Results}


\begin{table}[ht!]
    \centering
    \caption{Number of \bc{s} grouped by the effect of their arguments.}
    \label{tab:dabc-effects}
    \begin{tabular}{lrrr}
        \toprule
        \textbf{Category}    & \scikit & \pandas & \numpy \\
        \midrule
        \textit{Aesthetics}  & 1       & 5       & --     \\
        \textit{Behavior}    & 58      & 5       & 4      \\
        \textit{Performance} & 11      & 1       & 1      \\
        \textit{Refactoring} & 7       & --      & --     \\
        \bottomrule
    \end{tabular}
\end{table}

Table \ref{tab:dabc-effects} presents the number of \bc{s} classified in each category.
We observe that most \bc{s} performed in all libraries have an effect on the \textit{behaviour} of the function.
In total, 67 \bc{s} were categorized as such, 58 out of 77 (75\%) in \scikit, 5 out of 11 (45\%) in \pandas, and 4 out of 5 (80\%) in \numpy.
\textit{Performance}-based \bc{s} were also detected in all libraries, but at a lower number.
13 occurrences were classified in this category in total, 11 in \scikit\ (14\%), and one in both \pandas\ (9\%) and \numpy\ (20\%).

A few particularities have caught our attention as well.
First, 45\% of \pandas's \bc{s} were classified as \textit{Aesthetics}.
We suppose this number is high as aesthetics play an important role in \pandas\ since it provides functions to visualize the data managed by the library as well.
Second, only \scikit\ reported \bc{s} introduced as a consequence of \textit{refactoring} tasks; 7 out of 77 (9\%).
This result suggest that \scikit\ maintainers use refactoring as a way to keep compatibility with the library's new features and improvements.


\section{Discussion}
\label{sec:discussion}

We understand that the findings reported in this paper unfold implications for researchers, library maintainers, and library clients.
We discuss them in the following subsections.

\subsection{Implications for Researchers}

\paragraph{\bc{s} affect libraries in different ways}
As presented in Section \ref{sec:results-rqd}, third-party libraries can be affected in different severity levels by \bc{s}.
While 35\% of \scikit\ clients can be affected by at least one \bc{s}, only 0.13\% of \numpy\ clients are vulnerable.
Such disparity should be deeper investigated so we can better understand why \bc{s} have higher impact in some libraries.


\paragraph{We lack studies that investigate breaking changes in dynamically typed languages}
Public APIs frequently rely on function overloading to promote flexibility, readability, and reusability~\cite{Bloch2017, Lehman1979}.
Although dynamically-typed languages do not support this technique natively,\footnote{For more details see \url{https://softwareengineering.stackexchange.com/questions/425422/do-all-dynamically-typed-languages-not-support-function-overloading}} we can still make use of it---calling the same function with a variadic number of parameters---by using default argument values.
Consequently, we can say that languages like Python, JavaScript, and PHP are especially exposed to \bc{s}.
Therefore, we understand that dynamically typed languages should receive more attention in breaking changes studies~\cite{Lamothe2021, Mezzetti2018}, as more issues specific to these languages may arise.

\subsection{Implications for Library Maintainers}

\paragraph{\bc{s} present a ripple effect}
A few \bc{s} concentrate most of the issues for both maintainers and clients.
Results in Section \ref{sec:results-rqa} show that one argument---\mintinline{python}|cv|---is the pivot of 26\% of the \bc{s} we detected inside \scikit.
Moreover, Section \ref{sec:results-rqd} shows that three \bc{s} are responsible for more than 90\% of all occurrences in \pandas\ and \numpy\ client applications.
This finding suggests that, as with traditional breaking changes, \bc{s} may lead to a ripple effect among the API interface---updating the default value of one argument might affect the behaviour of other functions in the API---ultimately impacting a much larger number of client applications~\cite{Robbes2011, Ponomarenko2012}.
We argue that library maintainers should be aware of this issue when updating the default values of their APIs.

\paragraph{Maintainers should follow consistent versioning strategies to handle \bc{s}}
Results from Section \ref{sec:extended-analysis} reveal that most \bc{s} result in relevant behavior changes of their functions.
Furthermore, library maintainers adopt different procedures when introducing \bc{s}.
For instance, while \scikit\ added theirs only in major versions, \pandas\ included some in minors, and \numpy\ introduced some as a \textit{patch}.
Despite the particularities adopted for each library, we argue a more consistent versioning strategy for this kind of breaking change would benefit both clients and maintainers.

\subsection{Implications for Library Clients}

\paragraph{Clients should work with package managers}
The results reported in both Section \ref{sec:results-rqd} shows that client applications are vulnerable to \bc{s}; around 67K and 73K clients are exposed to \scikit\ and \pandas\ \bc{s}, respectively.
Moreover, clients turned out to be vulnerable to multiple API versions.
We reinforce a recommendation already mentioned in other works: client developers must adopt minimum versioning strategies when maintaining their dependencies, such as using package managers~\cite{Pimentel2019, Venturini2023, Mezzetti2018}.

\paragraph{Clients should choose carefully when to rely on default values}
As stated previously, using default argument values reduces the effort when using a given API method.
Section \ref{sec:background} shows that, although the \mintinline{python}|SVC| constructor receives 14 parameters, it is possible to create a new model without any specific as all arguments have default values assigned.
Under the hood, using default arguments introduces data dependencies in client applications as the values provided to their functions are defined by third parties, on which they have no control~\cite{Mezzetti2018, Ponomarenko2012}.
Some maintainers look aware of this issue, manifesting their concern during the introduction of some \bc{s}.
For instance, one \scikit\ maintainer stated that \textit{``people use default parameters when they shouldn't''}.\footnote{For more details, please check \url{https://github.com/scikit-learn/scikit-learn/issues/11128\#issue-326158504}.}
We argue that client developers should be diligent when relying on default values.
Specifically, we suggest temporarily relying on these values; for example, during development and experimentation stages.


\section{Threats to Validity}
\label{sec:threats}

\subsection*{Internal Validity}

\paragraph{Manual Analysis}
We answered some questions based on data manually analyzed by the authors.
For RQ.3, we leveraged and categorized the modules where the \bc{s} were located.
For RQ.5, we manually analyzed the comments of the issues that introduced the \bc{s} and pick the one that, in our view, better explained the reason.
For RQ.6, we read the description of the arguments responsible for the \bc{s} to understand which effect they could produce on clients.
These manual steps could introduce bias, as the authors' subjective judgment could influence the results.
We mitigated this risk by involving multiple authors to review and discuss the data from manual analysis

\subsection*{Construction Validity}

\paragraph{\bc{}s Identification Process}
We rely on the documentation provided by the libraries maintainers to leverage the \bc{s} investigated in this study, i.e.,~the \textit{versionchanged} attribute.
Naturally, this strategy may pose some threats in detecting API changes, as we are restricted to the changes properly documented in the library API.
In our favour, the maintainers of the three libraries follow strict guidelines for reporting API changes, providing clear instructions about how to document modifications in the library's API.\footnote{\url{https://scikit-learn.org/dev/developers/contributing.html\#change-the-default-value-of-a-parameter}}
Besides, other researchers also relied on API documentation to obtain breaking change candidates~\cite{Zhang2021}.

\paragraph{Function Call Heuristic}
We implemented a heuristic to identify \bc{s} calls in client applications.
Ideally, we could overcome this threat by executing the source code of each client and performing a dynamic analysis over the functions called.
Even though, we opted for a static-based analysis since previous works reported great difficulty in performing this task~\cite{Pimentel2019, Wang2020a}.
To ensure the reliability of our heuristic, we manually analyzed a sample of 384 calls and find out that $95.3\%$ ($\pm 5\%$) were correctly classified by the heuristic.

\subsection*{External Validity}

\paragraph{Selected Libraries}
We scoped this work on analyzing \bc{s} for the three most adopted Python libraries in Data Science.
Despite their popularity, analyzing only three popular Python data science libraries limits the generalizability of the findings.
The results may not apply to other libraries, ecosystems, or programming languages, thus threatening external validity.

\paragraph{Clients Dataset}
We rely on the dataset provided by Grotov et al.~\cite{Grotov2022} to investigate how clients are exposed to \bc{s}.
Although we could select other datasets to perform this analysis~\cite{Pimentel2019, Wang2020a, Quaranta2021, Zhang2021}, we take into account the documentation available to download and configure the dataset locally and the publicly available tool that---after proper adaptations---helped us in extracting the clients' function calls.
Yet, we understand it is important to expand this analysis to other artifacts besides Jupyter Notebooks, such as Python script files~\cite{Pimentel2019, Quaranta2021}.

\section{Related Work}
\label{sec:related-work}

\subsection{Library Updates and Breaking Changes}
Various studies proposed techniques to detect and understand breaking chan\-ges in libraries and frameworks~\cite{Mezzetti2018, Kula2018, Brito2020, Mostafa2017, Jezek2015, Dig2006}.
Kula et al.~\cite{Kula2018} conducted a study on the effect of API refactoring on client applications.
Their analysis involved versions of eight widely used libraries.
They discovered that many API-breaking changes were associated with bug fixes and new features.
Furthermore, 37\% of client-used API breakages were related to refactoring operations.
Mezzetti et al.~\cite{Mezzetti2018} conducted a study on breaking changes in the \textit{npm} repository and introduced a technique called \textit{type regression testing} to automatically detect whether a library update affects the types provided by its public interface. 
According to the authors, at least 5\% of all packages experienced a breaking change in a patch or minor update, with most of these changes attributed to modifications in the public package API.
Brito et al.~\cite{Brito2020} described an empirical study on popular Java libraries and frameworks where they continuously observed the changes performed in these libraries to understand why breaking changes are introduced.
The authors found that breaking changes were mostly motivated by the need to implement new features, simplify APIs, and improve maintainability.
They also analyzed 110 Stack Overflow posts and observed that breaking changes significantly impact client applications since 45\% of the questions ask about how to overcome specific breaking changes.
%
%
Mostafa et al.~\cite{Mostafa2017} describe a large-scale regression testing performed over 68 adjacent version pairs from 15 popular Java libraries to comprehend APIs' behavioural changes over time.
For this, the authors executed each version pair and compared the output produced by them, i.e.,~whether the updated code changed library behaviour.
Complementary, the authors also analyzed 126 real-world software bug reports about behavioural backward incompatibilities of software libraries.
Like us, the authors do not focus on signature incompatibilities that can be detected by compilers.
They investigate behaviour changes, i.e., for a given input, the changed code generates different outputs.
Their result reveals that behavioural backward-incompatibilities are common in Java libraries and are the root of many backward-incompatibility issues.

\subsection{APIs Smells in Data Science and Machine Learning}

The literature shows that both Data Science and Machine Learning fields are also prone to traditional software engineering issues~\cite{Sculley2015,Washizaki2022,Amershi2019, Gesi2022,Winters2020}.
For example, Tang et al.~\cite{Tang2021} study refactoring and technical debt issues in ML systems.
OBrien et al.~\cite{OBrien2022} also investigate technical debts in ML applications.
In the context of code smells, Zhang et al.~\cite{Zhang2022} identify 22 machine-learning-specific code smells, while Gesi et al.~\cite{Gesi2022} investigate code smell in Deep Learning software systems.
Regarding API maintenance, Haryono et al.~\cite{Haryono2021} studied a list of 112 deprecated APIs from three popular Python ML libraries to better understand how they can be migrated.
They aim to improve the comprehension of how updates could be applied to deprecated Machine Learning APIs.
For this, they collected a list of 112 APIs from three popular Python ML libraries: Scikit-Learn, TensorFlow, and PyTorch.
The authors identified three dimensions involving deprecated API migrations: update operation, API mapping, and context dependency.
Zhang et al.~\cite{Zhang2021} investigated changes performed on the API documentation of multiple TensorFlow versions to analyze their evolution.
Then, they classified these changes into ten categories according to the reason behind the modifications; the most common ones are efficiency and compatibility.

\begin{sidewaystable}[htbp]
    \small
    \centering
    \begin{tabularx}{1\linewidth}{c
            >{\hsize=0.8\hsize}X
            >{\hsize=0.8\hsize}X
            >{\hsize=1\hsize}X
            >{\hsize=1\hsize}X
            >{\hsize=1.4\hsize}X}
        \toprule
        \textbf{Study}      & \textbf{Components}                                          & \textbf{Type of Breaking Changes}      & \textbf{Goal}                                                        & \textbf{Scope}                                                                           & \textbf{Findings}                                                                                                          \\
        \hline
        \cite{Kula2018}     & Libraries and frameworks in Java, Python, and C\#            & Changes affecting APIs signature       & How clients are impacted by refactoring changes on libraries' APIs   & Eight libraries with 900 clients projects, approximately.                                & Library maintainers rarely break client-used APIs; Refactoring impacts less than 37\% of clients.                          \\ \hline
        \cite{Mezzetti2018} & Node.js libraries                                            & Changes of method signatures' types    & Assess if a library update changes the types in its public interface & Twelve libraries in Node.js ecosystem.                                                   & At least 5\% of all npm packages experience breaking changes in patch or minor updates.                                    \\ \hline
        \cite{Brito2020}    & Java libraries and frameworks                                & Both structural and behavioral changes & Understand why breaking changes are introduced                       & 400 real world Java libraries and frameworks and 110 posts on Stack Overflow.            & Breaking changes impact clients and are driven by the need to add new features, simplify APIs, and enhance maintainability \\ \hline
        \cite{Mostafa2017}  & Java libraries                                               & Behavioral changes                     & Understand behavioral changes in APIs during evolution of libraries  & 68 libraries version pairs and 126 bug reports on behavioral backward incompatibilities. & Behavioral backward incompatibilities are common in Java libraries and often cause significant compatibility issues.       \\ \hline
        \cite{Haryono2021}  & Machine learning frameworks (e.g., Scikit-Learn, TensorFlow) & Structural changes (i.e.,~Deprecation) & Understand how to update deprecated ML API usages in Python          & 112 APIs from Python ML libraries: Scikit-Learn, TensorFlow, and PyTorch.                & Three key dimensions were identified: Update Operation, API Mapping, and Context Dependency considerations                 \\ \hline
        \cite{Zhang2021}    & TensorFlow 2                                                 & Both structural and behavioral changes & Investigate how and why APIs evolve in deep learning frameworks      & 6,329 API changes across TensorFlow 2 versions.                                          & The most common changes are associated with efficiency and compatibility issues                                            \\ \bottomrule
    \end{tabularx}
    \caption{Comparison of our work with related studies.}
    \label{tab:related-work-comparisons}
\end{sidewaystable}

\subsection{Comparison with Related Work}

Table \ref{tab:related-work-comparisons} presents a comparison between our work and related studies regarding the libraries under analysis, the type of breaking changes investigated, the proposed goal and scope, and the main findings.
We decided to investigate breaking changes in Data Science libraries due to their increasing popularity and importance in the software development community;
other works share similar motivation~\cite{Haryono2021,Zhang2021}.
Unlike other structural changes~\cite{Kula2018, Brito2020, Haryono2021}, we focused on behaviour-breaking changes, as they are particularly difficult to spot in dynamically-typed languages~\cite{Mezzetti2018}.
Specifically, we investigated the impact of changes in default arguments since, for Machine Learning and Data Science applications, the use of this technique is very common.

\section{Conclusion}
\label{sec:conclusion}

In this work, we investigate an unexplored type of behaviour breaking change, which we named \textit{Default Argument Breaking Change (DABC)}.
We started by studying this phenomena in \scikit\ and analyzed how client applications are vulnerable to them~\cite{Montandon2023}.
Overall, we analyzed 77 \bc{s} among eight major versions of \scikit; 93\% of them were detected in client applications.

Next, we extended this work in two major directions.
First, we reproduced the initial study with two other libraries: \pandas\ and \numpy.
Second, we investigated the reason these \bc{s} were introduced according to the library maintainers and the effects caused by these changes in client applications.
A few \bc{s} are responsible for most vulnerabilities identified in the clients for all three libraries.
Besides, we observed that most \bc{s} were introduced to solve \textit{Maintainability} issues.
These \bc{s} modified their functions \textit{behavior} 87\% of the time (67 out 77 \bc{s}).
Finally, we discuss the implications of our findings for researchers, library maintainers, and clients.

\paragraph{Future Work}
We intend to extend this work in the following directions:
(a) measure the magnitude of the impact that \textit{behavior}-based \bc{s} may have on client applications;
(b) extend our analysis to other types of argument changes, such as their addition or removal;
(c) investigate the presence of \bc{s} in other ecosystems, such as JavaScript; and
(d) develop a tool to automatically detect the introduction of \bc{s} in library's source code.

\paragraph{Replication Package} Data and scripts are publicly available at Zenodo: \url{https://doi.org/10.5281/zenodo.11584961}.

\section*{Acknowledgment}

\noindent This research is supported by the National Council for Scientific and Technological Development (CNPq) and Fundação de Amparo à Pesquisa do Estado de Minas Gerais (FAPEMIG).

\bibliographystyle{elsarticle-num}
\bibliography{main}

\begin{thebibliography}{10}
\expandafter\ifx\csname url\endcsname\relax
  \def\url#1{\texttt{#1}}\fi
\expandafter\ifx\csname urlprefix\endcsname\relax\def\urlprefix{URL }\fi
\expandafter\ifx\csname href\endcsname\relax
  \def\href#1#2{#2} \def\path#1{#1}\fi

\bibitem{Provost2013}
F.~Provost, T.~Fawcett, Data Science for Business: What You Need to Know about Data Mining and Data-Analytic Thinking, 1st Edition, {O'Reilly}, {Beijing K\"oln}, 2013.

\bibitem{Ramasamy2022}
D.~Ramasamy, C.~Sarasua, A.~Bacchelli, A.~Bernstein, Workflow analysis of data science code in public {{GitHub}} repositories, Empirical Software Engineering 28~(1) (2022) 7.

\bibitem{Quaranta2021}
L.~Quaranta, F.~Calefato, F.~Lanubile, {{KGTorrent}}: {{A Dataset}} of {{Python Jupyter Notebooks}} from {{Kaggle}}, in: 18th {{International Conference}} on {{Mining Software Repositories}} ({{MSR}}), 2021, pp. 550--554.

\bibitem{Quaranta2022}
L.~Quaranta, F.~Calefato, F.~Lanubile, Eliciting {{Best Practices}} for {{Collaboration}} with {{Computational Notebooks}}, in: Proceedings of the {{ACM}} on {{Human-Computer Interaction}} ({{CSCW}}), Vol.~6, 2022, pp. 1--41.

\bibitem{NetflixTechnologyBlog2019}
{Netflix Technology Blog}, Beyond {{Interactive}}: {{Notebook Innovation}} at {{Netflix}}, https://netflixtechblog.com/notebook-innovation-591ee3221233 (Feb. 2019).

\bibitem{Winters2020}
T.~Winters, T.~Manshreck, H.~Wright, {Software Engineering at Google: Lessons Learned from Programming Over Time}, 1st Edition, O'Reilly Media, 2020.

\bibitem{Amershi2019}
S.~Amershi, A.~Begel, C.~Bird, R.~Deline, H.~Gall, E.~Kamar, N.~Nagappan, B.~Nushi, T.~Zimmermann, Software {{Engineering}} for {{Machine Learning}}: {{A Case Study}}, in: 41st {{ACM}}/{{IEEE International Conference}} on {{Software Engineering}} ({{ICSE}}), 2019.

\bibitem{Montandon2019b}
J.~E. Montandon, L.~Lourdes~Silva, M.~T. Valente, Identifying {{Experts}} in {{Software Libraries}} and {{Frameworks Among GitHub Users}}, in: 16th {{International Conference}} on {{Mining Software Repositories}} ({{MSR}}), 2019, pp. 276--287.

\bibitem{Haryono2021}
S.~A. Haryono, F.~Thung, D.~Lo, J.~Lawall, L.~Jiang, Characterization and {{Automatic Updates}} of {{Deprecated Machine-Learning API Usages}}, in: International {{Conference}} on {{Software Maintenance}} and {{Evolution}} ({{ICSME}}), 2021, pp. 137--147.

\bibitem{Grotov2022}
K.~Grotov, S.~Titov, V.~Sotnikov, Y.~Golubev, T.~Bryksin, A {{Large-Scale Comparison}} of {{Python Code}} in {{Jupyter Notebooks}} and {{Scripts}}, in: 19th {{International Conference}} on {{Mining Software Repositories}} ({{MSR}}), 2022, pp. 353--364.

\bibitem{Zhang2021}
Z.~Zhang, Y.~Yang, X.~Xia, D.~Lo, X.~Ren, J.~Grundy, Unveiling the {{Mystery}} of {{API Evolution}} in {{Deep Learning Frameworks}}: {{A Case Study}} of {{Tensorflow}} 2, in: 43rd {{International Conference}} on {{Software Engineering}}: {{Software Engineering}} in {{Practice}} ({{ICSE-SEIP}}), 2021, pp. 238--247.

\bibitem{Montandon2021}
J.~E. Montandon, M.~T. Valente, L.~L. Silva, Mining the {{Technical Roles}} of {{GitHub Users}}, Information and Software Technology 131 (2021) 1--19.

\bibitem{Montandon2021a}
J.~E. Montandon, C.~Politowski, L.~L. Silva, M.~T. Valente, F.~Petrillo, Y.~G. Gu{\'e}h{\'e}neuc, What skills do {{IT}} companies look for in new developers? {{A}} study with {{Stack Overflow}} jobs, Information and Software Technology 129~(August 2020) (2021) 1--6.

\bibitem{Brito2020}
A.~Brito, M.~T. Valente, L.~Xavier, A.~Hora, You broke my code: Understanding the motivations for breaking changes in {{APIs}}, Empirical Software Engineering 25 (2020) 1458--1492.

\bibitem{Ponomarenko2012}
A.~Ponomarenko, V.~Rubanov, Backward compatibility of software interfaces: {{Steps}} towards automatic verification, Programming and Computer Software 38 (2012) 257--267.

\bibitem{Venturini2023}
D.~Venturini, F.~R. Cogo, I.~Polato, M.~A. Gerosa, I.~S. Wiese, I depended on you and you broke me: {{An}} empirical study of manifesting breaking changes in client packages, ACM Transactions on Software Engineering and Methodology (2023).

\bibitem{Ochoa2022}
L.~Ochoa, T.~Degueule, J.-R. Falleri, J.~Vinju, Breaking bad? {{Semantic}} versioning and impact of breaking changes in {{Maven Central}}, Empirical Software Engineering 27 (2022) 1--42.

\bibitem{Mezzetti2018}
G.~Mezzetti, A.~M{\o}ller, M.~T. Torp, Type {{Regression Testing}} to {{Detect Breaking Changes}} in {{Node}}.js {{Libraries}}, in: 32nd {{European Conference}} on {{Object-Oriented Programming}} ({{ECOOP}}), 2018, pp. 1--24.

\bibitem{Ribeiro2016a}
M.~T. Ribeiro, S.~Singh, C.~Guestrin, "{{Why Should I Trust You}}?": {{Explaining}} the {{Predictions}} of {{Any Classifier}}, in: 22nd {{ACM International Conference}} on {{Knowledge Discovery}} and {{Data Mining}} ({{SIGKDD}}), 2016, pp. 1135--1144.

\bibitem{Pimentel2019}
J.~F. Pimentel, L.~Murta, V.~Braganholo, J.~Freire, A {{Large-Scale Study About Quality}} and {{Reproducibility}} of {{Jupyter Notebooks}}, in: 16th {{International Conference}} on {{Mining Software Repositories}} ({{MSR}}), 2019, pp. 507--517.

\bibitem{Wang2020a}
J.~Wang, T.-Y. KUO, L.~Li, A.~Zeller, Assessing and {{Restoring Reproducibility}} of {{Jupyter Notebooks}}, in: 35th {{International Conference}} on {{Automated Software Engineering}} ({{ASE}}), 2020, pp. 138--149.

\bibitem{Montandon2023}
J.~E. Montandon, L.~L. Silva, C.~Politowski, G.~E. Boussaidi, M.~T. Valente, Unboxing {{Default Argument Breaking Changes}} in {{Scikit Learn}}, in: 23rd {{IEEE International Working Conference}} on {{Source Code Analysis}} and {{Manipulation}} ({{SCAM}}), 2023, pp. 1--11.

\bibitem{Phillips2018}
D.~Phillips, Python 3 Object-Oriented Programming: Build robust and maintainable software with object-oriented design patterns in Python 3.8, 3rd Edition, Packt Publishing, 2018.

\bibitem{Bloch2017}
J.~Bloch, Effective {{Java}}, 3rd Edition, {Addison-Wesley Professional}, 2017.

\bibitem{Lehman1979}
M.~M. Lehman, On understanding laws, evolution, and conservation in the large-program life cycle, Journal of Systems and Software 1 (1979) 213--221.

\bibitem{Pedregosa2011}
F.~Pedregosa, G.~Varoquaux, A.~Gramfort, V.~Michel, B.~Thirion, O.~Grisel, M.~Blondel, P.~Prettenhofer, R.~Weiss, V.~Dubourg, J.~Vanderplas, A.~Passos, D.~Cournapeau, M.~Brucher, M.~Perrot, {\'E}.~Duchesnay, Scikit-learn: {{Machine Learning}} in {{Python}}, Journal of Machine Learning Research 12~(85) (2011) 2825--2830.

\bibitem{Wang2016}
D.~Wang, P.~Cui, W.~Zhu, Structural {{Deep Network Embedding}}, in: 22nd {{International Conference}} on {{Knowledge Discovery}} and {{Data Mining}} ({{KDD}}), 2016, pp. 1225--1234.

\bibitem{Bianchi2021}
F.~Bianchi, S.~Terragni, D.~Hovy, Pre-training is a {{Hot Topic}}: {{Contextualized Document Embeddings Improve Topic Coherence}}, in: 59th {{Annual Meeting}} of the {{Association}} for {{Computational Linguistics}} and the 11th {{International Joint Conference}} on {{Natural Language Processing}} ({{IJCNLP}}), 2021, pp. 759--766.

\bibitem{Domingos2012}
P.~Domingos, A few useful things to know about machine learning, Commun. ACM 55~(10) (2012) 78--87.

\bibitem{Tan2020}
J.~Tan, D.~Feitosa, P.~Avgeriou, Investigating the {{Relationship}} between {{Co-occurring Technical Debt}} in {{Python}}, in: 46th {{Euromicro Conference}} on {{Software Engineering}} and {{Advanced Applications}} ({{SEAA}}), 2020, pp. 487--494.

\bibitem{Hinkle2003}
D.~E. Hinkle, W.~Wiersma, S.~G. Jurs, Applied Statistics for the Behavioral Sciences, 5th Edition, {Wadsworth Cengage Learning}, {Belmont, CA}, 2003.

\bibitem{Borges2018}
H.~Borges, M.~Tulio~Valente, What's in a {{GitHub Star}}? {{Understanding Repository Starring Practices}} in a {{Social Coding Platform}}, Journal of Systems and Software 146 (2018) 112--129.

\bibitem{Spencer2009}
D.~Spencer, J.~J. Garrett, Card Sorting: Designing Usable Categories, {Rosenfeld Media}, 2009.

\bibitem{Warrens2015}
M.~J. Warrens, Five {{Ways}} to {{Look}} at {{Cohen}}'s {{Kappa}}, Journal of Psychology \& Psychotherapy 05~(04) (2015).

\bibitem{Lamothe2021}
M.~Lamothe, Y.-G. Gu{\'e}h{\'e}neuc, W.~Shang, A {{Systematic Review}} of {{API Evolution Literature}}, ACM Computing Surveys 54~(8) (2021) 171:1--171:36.

\bibitem{Robbes2011}
R.~Robbes, M.~Lungu, {A Study of Ripple Effects in Software Ecosystems}, in: 33rd International Conference on Software Engineering (ICSE, NIER Track), 2011, pp. 904--907.

\bibitem{Kula2018}
R.~G. Kula, A.~Ouni, D.~M. German, K.~Inoue, An empirical study on the impact of refactoring activities on evolving client-used apis, Information and Software Technology 93~(C) (2018) 186--199.

\bibitem{Mostafa2017}
S.~Mostafa, R.~Rodriguez, X.~Wang, Experience paper: A study on behavioral backward incompatibilities of {{Java}} software libraries, in: 26th {{International Symposium}} on {{Software Testing}} and {{Analysis}} ({{ISSTA}}), 2017, pp. 215--225.

\bibitem{Jezek2015}
K.~Jezek, J.~Dietrich, P.~Brada, How java apis break - an empirical study, Information and Software Technology 65~(C) (2015) 129--146.

\bibitem{Dig2006}
D.~Dig, R.~Johnson, How do apis evolve? a story of refactoring: Research articles, Journal of Software Maintenance and Evolution 18~(2) (2006) 83--107.

\bibitem{Sculley2015}
D.~Sculley, G.~Holt, D.~Golovin, E.~Davydov, T.~Phillips, D.~Ebner, V.~Chaudhary, M.~Young, J.-F. Crespo, D.~Dennison, Hidden {{Technical Debt}} in {{Machine Learning Systems}}, in: 28th {{International Conference}} on {{Neural Information Processing Systems}} ({{NIPS}}), 2015.

\bibitem{Washizaki2022}
H.~Washizaki, F.~Khomh, Y.-G. Gu{\'e}h{\'e}neuc, H.~Takeuchi, N.~Natori, T.~Doi, S.~Okuda, Software-{{Engineering Design Patterns}} for {{Machine Learning Applications}}, Computer 55~(3) (2022) 30--39.

\bibitem{Gesi2022}
J.~Gesi, S.~Liu, J.~Li, I.~Ahmed, N.~Nagappan, D.~Lo, E.~S. de~Almeida, P.~S. Kochhar, L.~Bao, Code smells in machine learning systems (2022).

\bibitem{Tang2021}
Y.~Tang, R.~Khatchadourian, M.~Bagherzadeh, R.~Singh, A.~Stewart, A.~Raja, An empirical study of refactorings and technical debt in machine learning systems, in: IEEE/ACM 43rd International Conference on Software Engineering (ICSE), 2021, pp. 238--250.

\bibitem{OBrien2022}
D.~OBrien, S.~Biswas, S.~M. Imtiaz, R.~Abdalkareem, E.~Shihab, H.~Rajan, 23 {{Shades}} of {{Self-Admitted Technical Debt}}: {{An Empirical Study}} on {{Machine Learning Software}}, in: 30th {{ACM Joint European Software Engineering Conference}} and {{Symposium}} on the {{Foundations}} of {{Software Engineering}} ({{ESEC}}/{{FSE}}), 2022, p.~13.

\bibitem{Zhang2022}
H.~Zhang, L.~Cruz, A.~van Deursen, Code smells for machine learning applications, in: 1st International Conference on AI Engineering: Software Engineering for AI, CAIN'22, 2022, pp. 217--228.

\end{thebibliography}

\end{document}